\documentclass[11pt,a4]{article}

\usepackage{amsmath,amsthm}
\usepackage{graphicx}
\usepackage{amsfonts}
\usepackage{color,multirow,hyperref}
\usepackage{epstopdf}
\usepackage{authblk}
\usepackage[margin=3cm]{geometry}
\usepackage{rotating}

\newcommand{\R}{\mathbb{R}}
\def\reffig#1{Fig.~\ref{#1}}

\theoremstyle{definition}
\newtheorem{remark}{Remark}

\providecommand{\keywords}[1]
{
  \small	
  \textbf{\textit{Keywords---}} #1
}

\date{} 

\title{Free-form Design of Discrete Architectural Surfaces \\ by use of Circle Packing}

\author[$1$]{Shizuo Kaji}
\affil[$1$]{Institute of Mathematics for Industry, Kyushu University, Japan}

\author[$2$]{Jingyao Zhang
\thanks{corresponding author}}
\affil[$2$]{Dept. of Architecture and Architectural Engineering, Kyoto University, Japan}

\begin{document}

\maketitle

\begin{abstract}
	This paper presents an efficient approach for the conceptual design of architectural surfaces
	which are composed of triangular panels.
    {In the free-form design of discrete architectural surfaces,
    the Gaussian curvature plays an important role not only aesthetically but also
    in terms of stiffness and constructability. 
    However, designing a surface manually with specific Gaussian curvatures can be a time-consuming task.
	We propose a method to find a triangulated surface with user-specified Gaussian curvatures (not limited to constant Gaussian curvatures) and boundary vertex positions.}
	In addition, the \emph{conformal class} of the final design can be specified; that is, the user has  control over the shape (the corner angles) of each triangular panel. 
	The panels could be encouraged to form a regular tessellation or kept close to those of the initial design.
	The controllability of the conformal class suppresses possible distortion of the panels,
	resulting in higher structural performance and aesthetics.
	
	Our method relies on the idea in {computational conformal geometry} called \emph{circle packing}.
	In this line of research, the discrete Ricci flow has been widely used for surface modelling.
	However, it is not trivial to incorporate constraints such as boundary locations and convexity of the spanned surface, which are essential to architectural applications. 
	We propose a perturbation of the discrete Ricci energy and develop a least-squares-based optimisation scheme to address these problems with an open-source implementation\footnote{Our codes are publicly available at \url{https://github.com/shizuo-kaji/ricci_flow}}.
\end{abstract}



\section{Introduction}
Upon the rapid development in digital technology,
free-form surfaces have found their progressively important position in many fields such as
product design, ship building, and architecture.
Tools have been developed for the creation of digital models,
where surfaces designed by artists are algorithmically modified to meet practical and aesthetic requirements
while retaining the qualities of the original surfaces including the artists' design intent (e.g., \cite{jcde-qwab036}).
In the field of architecture, \emph{architectural geometry}~\cite{Pottmann2010}
aims to solve 
complex and challenging problems in designing
architectural surfaces under various geometric and structural constraints
by combining knowledge from
differential geometry, computational geometry, and architectural design and engineering.

Most of the modern architectural surfaces are intrinsically discrete,
with only a few exceptions that are truly continuous such as
membrane and concrete shell structures.
{The geometry of discrete surfaces has been extensively studied 
in the field of \emph{discrete differential geometry}~\cite{Crane2020}
and computational conformal geometry~\cite{CCG_2007},
One of the pioneering work utilising computational conformal geometry in architecture
is \cite{Schiftne2009}, where
the idea of circle packing is used  to create 
triangular meshes whose incircles form a circle packing on the surface.
Another prominent example is \cite{Quasiisothermic}, which develops approximation 
of surfaces by curvature-aligned rectangular panels.}

{
A prevalent idea employed by many geometric methods is to formulate 
the problem of finding surfaces with specified design targets 
as variational problems with certain energy functionals.
Among the design targets involving 
geometric property of a surface, the Gaussian curvature serves one of the most important indices.
This preference relies not only on \emph{aesthetics} but also on \emph{stability} (stiffness), 
which are two of the three design principles\footnote{The third design principle is \emph{functionality}.}
advocated by Vitruvius (15 B.C.) and accepted by most prominent architects.
In the aesthetic aspect, curved lines and surfaces are more harmonious, relaxing, and pleasant,
because human preference for curvature is biologically determined~\cite{Gomez-Puerto2016}.
The architectural surface or facade is of course not an exception as investigated in~\cite{Ruta2019}.
In the stability aspect, 
larger curvature generally leads to higher stiffness.
For instance, the Guggenheim Museum makes use of  the ever-present curvature of the various geometries,
not only for aesthetic purpose but also to significantly enhance stiffness of the structure against lateral loads~\cite{Novak1998}.
}
In the past decades, double curvature structural envelopes
(that is, non-developable surfaces) have become increasingly popular in architectural design.

{
It is usually desirable to have gradually varying curvature, if not constant, over the surface,
in order to present its elegant appearance and to avoid sudden change in stiffness distribution.
}
It is, however, challenging to design surfaces manually that have
specified Gaussian curvatures within a marginal error allowed by constructability requirements.
One designing principle is to specify the boundary of the surface
and a rough (expected) shape that spans the boundary, and rely on an algorithmic technique to find the final surface numerically.
For example, a numerical form-finding method was presented based on the generation of linear Weingarten surfaces~\cite{Tellier2019}.
This method allows to build the surfaces on a target boundary curve
but does not allow the user to specify the Gaussian curvatures of the interior vertices explicitly.
Another method is proposed to obtain a surface with user-specified Gaussian curvatures
based on the Gaussian curvature flow~\cite{Zhao2006}.
However, it is shown to be highly unstable.
Furthermore, the shape of the panels could be distorted in the obtained surface even when the initial surface consists of near-regular panels.
Keeping (regular) shapes of the panels is important
to aesthetic appearance of the surface and to structural performance.
To preserve the angles of the panels of the initial surface,
Ricci flow-based\footnote{{In dimension two, the Ricci flow coincides with the Yamabe flow and
these names are used interchangeably.}} methods have been developed~\cite{Jin2008,Yang2009,Zhang2014f,Jin2018}, which find final surfaces \emph{conformal} to the initial one (see \cite{Zeng2013} for a survey).
However, existing applications of Ricci flow-based algorithms mainly focus on flattening surfaces to have
the constant Gaussian curvatures at all vertices {(e.g., \cite{springborn})}.
This agrees with the fact that the Ricci flow is closely related to the uniformisation theorem~\cite{Hamilton1988}.
Furthermore, existing Ricci flow-based algorithms do not pay much attention 
to imposing constraints such as boundary locations and convexity (or concavity) of the resulting surface, which are relevant to architectural applications.
{A mesh editing technique developed in \cite{Vaxman2015}
finds a near-conformal mesh satisfying positional constraints by minimising
an energy functional measuring the deviation of two meshes from conformally equivalent.
However, it is not straightforward to prescribe curvatures with this method.}

These problems in designing discrete architectural surfaces
motivate us to present an efficient tool for quick conceptual design,
which ``finetune'' the initial surface to one 
with user-specified Gaussian curvatures and boundary vertex locations within 
a user-specified conformal class.
The user can focus on the rough design of the surface, and our system 
turns the surface automatically to one that meets curvature, boundary, and panel shape requirements.
Our flexible framework allows the user to incorporate other constraints as well.
As an example, finding a convex surface with user-specified Gaussian curvatures is demonstrated.

After the introduction, this paper is organised as follows:
Section \ref{sec:overview} describes the background of discrete surfaces and circle packing on them.
Section \ref{sec:algorithm} discusses the algorithmic details of our method.
Section \ref{sec:example} presents numerical examples to validate the proposed approach and
	demonstrate its versatility for the free-form design of discrete surfaces.
Section \ref{sec:conclusion} concludes the study with limitations of the proposed method and visions of future works.

\section{Overview of discrete conformal surface geometry}\label{sec:overview}

This section provides an overview of the surface modelling based on computational conformal geometry.
We put a particular emphasis on various geometric structures on discrete surfaces and their hierarchy.
Although we often use the terminology in differential geometry to evoke analogy, 
we do not rely on the knowledge of differential geometry, and we give a definition to each notion relevant in our setting.

\subsection{Triangulated surface mesh}
In this study, we consider surfaces that are triangulated meshes.
{
Triangulated mesh is one of the most popular surface representation with which
various mesh processing algorithms have been developed. 
As our method takes and outputs triangulated meshes,
it can be incorporated as a module of the surface design toolchain.
There exist algorithms to convert triangular meshes into rectangular ones 
such as the Catmull–Clark subdivision~\cite{CC} and \cite{Quasiisothermic}, which can be applied as the post process of our method.
From an architectural perspective,
intrinsic mechanical stability of triangles
leads to much higher stiffness-to-material ratio compared to other polygonal shapes.
Their superiority in lightweightness or material-efficiency enables the 
triangular meshes to be suitable for free-form architectures.
This is in particular the case for the roof structures that span a huge column-free space,
when they are utilized as supporting system themselves~\cite{Meza2021}.
The Beijing Daxing International Airport Terminal, jointly designed by the Zaha Hadid Architects,
is an excellent example:
a grid shell with triangular mesh is used for the free-form roof of the center zone.
Due to this unique roof structure, together with other innovations, 
e.g., the biggest terminal and the largest vibration-isolation building in the world~\cite{Zhou2020},
it has won a number of awards and was named as one of the new seven wonders by The Guardian. 
}


An abstract triangulated mesh (a mesh, for short) is composed of the set $\mathcal{V}$ of  vertices, the set $\mathcal{E}$ of
edges, and the set $\mathcal{F}$ of triangular faces.
Here, we have architectural surfaces composed of triangular panels (faces) in mind.
The edge connecting two vertices $i,j\in \mathcal{V}$ is denoted by $e_{ij}$,
and the face formed by three vertices $i$, $j$, and $k$ is denoted by $f_{ijk}$.
As we are modelling surfaces, we always assume that 
the neighbourhood of each vertex is homeomorphic to the plane or the half-plane,
and each edge is contained in at most two faces.
An edge is said to be \emph{interior} (resp. \emph{boundary})
if it is contained in two faces (resp. a single face).
The set of boundary edges is denoted by $\mathcal{E}_{\textrm{bd}}$.
%
The vertices contained in boundary edges are called \emph{boundary vertices},
and the set of boundary vertices is denoted by $\mathcal{V}_{\textrm{bd}}$.

\begin{figure*}[htb]
	\centering
    	\includegraphics[width=0.2\textwidth]{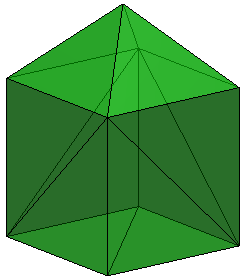} \hspace{1cm} \includegraphics[width=0.2\textwidth]{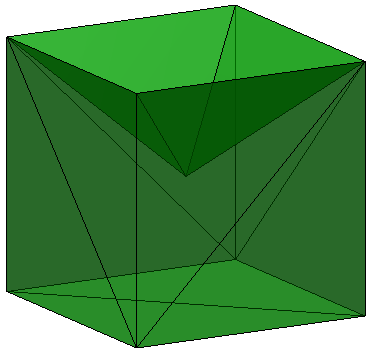} 
	\caption{Example of different embeddings of a mesh with a metric.
	These two shapes consist of exactly the same set of panels with the same connection
	among them.}
	\label{Fig:house}
\end{figure*}

A mesh is a combinatorial object and retains only the topology of the surface; that is, no information on lengths, areas, vertex positions are associated.
The information on sizes can be bestowed by edge lengths $\ell: \mathcal{E}\to \R_{\ge 0}$; an assignment of non-negative number $\ell(e_{ij})=l_{ij}$ for each edge $e_{ij}$. 
Since faces are triangles, we demand the edge lengths to satisfy
the triangular inequality $l_{ij} + l_{jk} > l_{ik}$ for any face $f_{ijk}$.
In this case, we say the edge lengths $\{l_{ij}\}$ define a \emph{discrete Riemannian metric} (a metric, for short) on a mesh.
Note that the edge lengths completely determine the shapes of the triangle faces,
but not the dihedral angles between them;
one may build more than one surfaces with the same set of panels
(see \reffig{Fig:house}).

The Gaussian curvature at a vertex quantifies the deviation of 
the neighbourhood of the vertex from a plane.
There are several (non-equivalent) definitions of Gaussian curvatures 
for a mesh with a metric.
In this paper, we adopt the angle defect as their definition.
Namely, the Gaussian curvature $K_i$ at vertex $i$ is defined as follows:
\begin{align}
	\label{Eq:curvature}
		K_i &= 2\pi - \sum_{{f}_{ijk}\in \mathcal{F}}\theta_{i}^{jk}
		\qquad
		(i\not\in \mathcal{V}_{\textrm{bd}}), \nonumber\\
		K_i &= \pi - \sum_{{f}_{ijk}\in \mathcal{F}}\theta_{i}^{jk}
		\qquad
		(i\in \mathcal{V}_{\textrm{bd}}),
\end{align}
where $\theta_i^{jk}$ is the corner angle at vertex $i$ in the face ${f}_{ijk}$.
{The Gaussian curvatures are constantly zero for any triangulation of the flat plane 
and surfaces foldable from it (imagine edges work as hinges).}
The Gaussian curvatures of a mesh are computed solely from its metric
independent of the coordinates of its vertices; 
such a geometric property is said to be \emph{intrinsic}.
{That is, intrinsic properties of a surface are determined by the shape of its constituent panels
and how they are connected each other regardless of the location (coordinates) of the panels.
Thus, intrinsic properties are preserved when the initial surface is folded to another one without changing the shape of its faces.}

Although a metric determines the local shape of a mesh,
its global realisation as a surface in space is not specified merely by a metric.
An \emph{embedding} $\psi: \mathcal{V}\to \R^3$ of a mesh assigns 
the coordinates $v_i=\psi(i)\in \R^3$ to each vertex $i\in \mathcal{V}$.
\reffig{Fig:house} shows an example of two embeddings
of a mesh with a metric; that is, two surfaces constructed from the same panels with the same connection among them.
They have the same edge lengths, and hence, the same Gaussian curvatures.
We call a mesh together with an embedding \emph{design} or \emph{geometry}.
{In summary, a design is specified by or decomposed into three components; 
a mesh (connection of panels), a metric (shape of panels), and an embedding (location of panels).}

\subsection{Circle packing on a mesh}

\begin{figure*}[htb]
	\centering
    \begin{tabular}{ccc}
    	\includegraphics[width=0.3\textwidth]{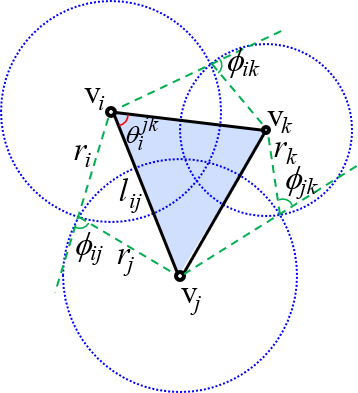} & $\quad\quad$ &
    	\includegraphics[width=0.3\textwidth]{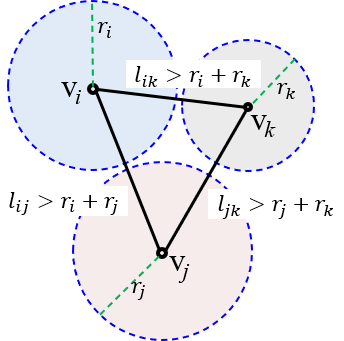} \\
    	(a) Thurston's circle packing & &
    	(b) general circle packing
    \end{tabular}
	\caption{Circle packing on a mesh.
	The conformal structure $\eta_{ij}$ generalises the cosine of the intersection
	angle of adjacent circles using the cosine law.}
	\label{Fig:circle}
\end{figure*}

For smooth surfaces, conformal deformations are characterised as
deformations that preserve angles.
However, a general discrete surface represented by a triangulated mesh
cannot be deformed to a non-similar one
without changing the corner angles of the triangles.
Obviously, we need more flexibility to design architectural surfaces.
To avoid this rigidity (or locking) problem, we need a different notion of 
conformality in the discrete setting. 
There are several non-equivalent definitions of conformality of discrete surfaces (see, e.g., \cite{crane13,Bobenko2016}),
we rely on Thurston's idea \cite{Thurston1980} to look at \emph{circle packing}.
Under a conformal deformation of a smooth surface, 
the intersection angles of infinitesimal circles covering the surface do not change. 
In analogy, for a triangular mesh, we consider a circle packing
in which a circle of finite radius $r_i> 0$ is assigned to each vertex $i$.
Colloquially, a deformation is conformal if it does not alter the intersection angles
of the circles (this analogy is made precise in~\cite{Rodin1987}).
The intersection angle (see \reffig{Fig:circle}(a))
$\phi_{ij}$ between the circles at vertices $i$ and $j$
is computed by:
\begin{align}
	\label{Eq:circle2}
	\cos\phi_{ij} = \frac{l_{ij}^{2} - r_i^2 - r_j^2 }{2r_ir_j},
\end{align}
where $i$ and $j$ are connected by an edge $e_{ij}$.
\emph{Thurston's circle packing} of a mesh additionally assumes
that the circles intersect with non-obtuse angles $\{\phi_{ij}\in [0,\pi/2]\}$.
However, it is not easy to find Thurston's circle packing in practice for a given mesh with a metric.
Hence, we rely on a relaxed and abstract notion of \emph{conformal structure} on a mesh,
which is an assignment of $\eta_{ij}\ge 0$ for each edge $e_{ij}$.
For a Thurston circle packing, we define $\eta_{ij}=\cos\phi_{ij}$.
We also allow ``non-intersecting'' circle packing, which is called the inversive distance circle packing~\cite{Bowers2004},
as in \reffig{Fig:circle}(b).

For a mesh with a circle packing $\{r_i\}$ and a conformal structure $\{\eta_{ij}\}$, 
a metric $\{l_{ij}\}$ can be associated by
\begin{align}
	\label{Eq:circle1}
	l_{ij}^2 = r_i^2 + r_j^2 + 2 r_ir_j\eta_{ij} \quad (\forall e_{ij}\in \mathcal{E}).
\end{align}
From this relation, a conformal structure together with a circle packing determines a metric.
By rewriting the formula,
we see a metric together with a circle packing determine a conformal structure analogiously to
\eqref{Eq:circle2}:
\begin{align}
	\label{Eq:eta}
	 \eta_{ij} = \dfrac{l_{ij}^2 - r_i^2 - r_j^2}{2 r_ir_j}.
\end{align}
A deformation that preserves the discrete conformal structure $\{\eta_{ij}\}$
is called a \emph{conformal deformation}.

For a given mesh with a metric, 
there are several heuristic ways to define a circle packing radius $r_i$ at vertex $i$
from the metric.
A popular choice \cite{Bowers2004} is given by
\begin{align}
	\label{Eq:inversive}
	r_i = \frac{1}{2}\min_{f_{ijk}\in \mathcal{F}}(l_{ik} + l_{ij} - l_{jk}).
\end{align}
Alternatively, we can ignore the metric, and set $\eta_{ij}=1$ for all $e_{ij}\in \mathcal{E}$,
which would correspond to the conformal structure of a regular tessellation.

\subsection{Constraint on Gaussian curvature}
The primary goal of this paper is to find an embedding of a mesh that achieves
user-specified Gaussian curvatures. We would like to recall the fact that
the Gaussian curvatures cannot take arbitrary values 
but are constrained by the mesh topology and geometry.
First, the Gauss-Bonnet theorem imposes the topological constraint:
\begin{align}
	\label{Eq:gauss-bonnet}
	\sum_{i \in \mathcal{V}} K_i = 2\pi \chi(\mathcal{F}),
\end{align}
where $\chi(\mathcal{F}) = |\mathcal{V}| - |\mathcal{E}| + |\mathcal{F}|$
is the Euler characteristic of the mesh.
Second, the conformal structure imposes the geometric constraint~\cite{Chow2003a,Jin2008}:
For any proper subset $I\subset \mathcal{V}$,
\begin{align}
	\label{Eq:admissible}
    \sum_{i\in I}K_i > -\sum_{(e_{jk},i)\in Lk(I)}(\pi-\eta_{jk})+2\pi\chi(F_I),	
\end{align}
where $F_I$ is the set of the faces whose vertices are in $I$,
and $Lk(I)=\{(e_{jk},i)\in \mathcal{F}\mid i\in I, \ j,k\not\in I\}$ is the link of $I$.

\begin{remark}
{With dynamic Yamabe flow algorithms which allow modification of the mesh topology, 
the condition \eqref{Eq:admissible} is not necessary (see \cite{sun2015discrete,Gu_JDG_1}}).
However, we stick to our constraint of preserving the initial mesh topology
so that, for example, we can aim at a near regular mesh.
\end{remark}

\subsection{Metric deformation of a mesh}
As a Riemannian metric determines the Gaussian curvatures of a smooth surface, a metric on a mesh (edge lengths) determines the corner angles of the triangular faces by cosine law, and in turn, the Gaussian curvatures of its vertices.
To find a metric having the user-specified Gaussian curvatures, 
the Ricci flow can be utilised.
The Ricci flow for discrete surfaces was established by \cite{Chow2003a}
in analogy to its smooth counterpart introduced in~\cite{Hamilton1988}. 
It was shown that under a mild condition, the discrete Ricci flow converges quickly to give a metric 
having the specified Gaussian curvatures while preserving the conformal structure.
As our method is closely related to the discrete Ricci flow, we give a brief description of it.
We refer the reader to \cite{Zeng2013} for details of the discrete Ricci flow.

With a fixed mesh with a fixed discrete conformal structure, 
we are to find such a metric 
that gives a Gaussian curvature close to a user-specified one $\bar{K}_i$ at each vertex $v_i$.
The key point is that we deform the circle radii 
instead of directly deforming the metric or the vertex positions.
Recall that with a fixed conformal structure $\{\eta_{ij}\}$,
the metric $\{l_{ij}\}$ is reconstructed by 
the circle radii $\{r_i\}$ by \eqref{Eq:circle1}.
The strategy of deforming the circle radii instead of the edge lengths has two advantages. 
First, one can specify the conformal class of the metric.
Second, edge lengths obtained by \eqref{Eq:circle1} automatically satisfy triangular inequality 
so that we do not have to care the space of feasible solutions.

Denote the vector $(u_1,u_2,\ldots,u_{|\mathcal{V}|})$ by ${\bf u}$,
where $u_i = \ln {r}_i$.
Given a prescribed Gaussian curvature $\bar{K}_i$ for each
vertex $i\in \mathcal{V}$, the Ricci energy is defined by
\begin{align}
	\label{Eq:ricci}
	E({\bf u}) &= \int_{\mathbf{0}}^{\bf u} \sum_{i\in \mathcal{V}} (K_i(\mathbf{u}) - \bar{K}_i) {\rm d}u_i \nonumber\\
	&=
	\int_0^{\mathbf{u}} \sum_{i\in \mathcal{V}} K_i(\mathbf{u})  {\rm d}u_i - \sum_{i\in \mathcal{V}} \bar{K}_i u_i,
\end{align}
where $\mathbf{0}=(0,0,\ldots,0)$ is the null vector.
The integral does not depend on the path as the integrand can be shown to be an exact form.
It is proved in \cite{Chow2003a} that 
when $\{\bar{K}_i\}$ satisfy \eqref{Eq:admissible},
the Ricci energy is strictly convex on the space
 $\sum_{i\in \mathcal{V}} u_i = 0$, which amounts to fixing the indeterminacy of the uniform scaling.
The discrete Ricci flow is the negative gradient flow of the Ricci energy, which is calculated as
\begin{align}
	\label{Eq:gradient}
	\dfrac{d \mathbf{u}}{dt} &= -\nabla E({\mathbf{u}})\nonumber\\
	&=  ( \bar{K}_1 - K_1(\mathbf{u}) , \bar{K}_2 - K_2(\mathbf{u}), \nonumber\\
	  &\quad \quad \cdots, \bar{K}_{|\mathcal{V}|} - K_{|\mathcal{V}|}(\mathbf{u}) ).
\end{align}
Under the conditions \eqref{Eq:gauss-bonnet} and \eqref{Eq:admissible},
the Ricci energy attains its global minimum when $\nabla E(\mathbf{u})=0$, which means
$K_i(\mathbf{u})=\bar{K}_i$ for all $i\in \mathcal{V}$.

While the Ricci flow provides a powerful method for mesh deformation backed with a theoretical guarantee, 
there are some difficulties when one tries to apply it for practical problems.
While the gradient of the Ricci energy has a simple closed-form expression \eqref{Eq:gradient},
its value itself is defined by an integral \eqref{Eq:ricci}.
This hinders the use of most of modern optimisation techniques such as trust region which require the evaluation of the target function value.
Our idea is to modify the Ricci energy so that it admits a closed, least-squares form,
which allows us to use well-established solvers and to incorporate various constraints 
such as boundary positions and convexity. 
We propose the modified Ricci energy defined by
\begin{align}
	\label{Eq:modifiedEnergy}
	\hat{E}(\mathbf{u}) = \sum_{i\in \mathcal{V}_K} (K_i(\mathbf{u})-\bar{K}_i)^2,
\end{align}
where $\mathcal{V}_K\subset \mathcal{V}$ is a subset of vertices on which we would like to specify their Gaussian curvatures 
(typically, the set of interior vertices is chosen).
The above mentioned properties of the Ricci energy guarantee that
there exists a unique $\mathbf{u}$ which realises the global minimum
$\hat{E}(\mathbf{u})=0$ when $\mathcal{V}_K=\mathcal{V}$ and $\{\bar{K}_i\}$ satisfy \eqref{Eq:gauss-bonnet} and \eqref{Eq:admissible}.
We cannot, however, assert that there are no other local minima.
The gradient of \eqref{Eq:modifiedEnergy} is easily computed as
\[
\dfrac{\partial \hat{E}}{\partial u_j}(\mathbf{u}) = 
2\sum_{i\in \mathcal{V}_K} (K_i(\mathbf{u})-\bar{K}_i)\dfrac{\partial K_i}{\partial u_j}(\mathbf{u}),
\]
where $\dfrac{\partial K_i}{\partial u_j}$
is given, e.g., in \cite{Jin2008}.

Once the circle radius ${r}_i=\exp{u_i}$ for each vertex ${v}_i$ is obtained by minimising \eqref{Eq:modifiedEnergy},
the edge lengths ${l}_{ij}$ are computed by \eqref{Eq:circle1},
while the discrete conformal structure $\{\eta_{ij}\}$ is kept unchanged from their initial values.


\section{Implementation of our method}\label{sec:algorithm}

In the previous section, we summarised how various properties are determined on the surface as functions of other properties.
A mesh defines the topology and serves as a container of geometric structures.
{It carries the information on the number of triangle panels and how they are connected.
A metric is a function $\ell: \mathcal{E}\to \R$ specifying
the edge lengths $\{l_{ij}\in \R|e_{ij}\in \mathcal{E}\}$,
which determines the shape of each panel.
An embedding is a function $\psi: \mathcal{V}\to \R^3$ specifying the vertex coordinates
$\{v_i\in \R^3|i\in \mathcal{V}\}$, that is, how each panel is placed.
The Gaussian curvatures are solely determined by the metric (the panel shape) independent of an embedding.
Therefore, to obtain a surface with prescribed Gaussian curvatures,
we have to find the appropriate metric.
Instead of handling the edge lengths directly, we can consider proxy variables.
A circle packing is a function $r: \mathcal{V}\to \R$ specifying the radius of a circle centred at each vertex.
A conformal structure $\eta: \mathcal{E}\to \R$ together with a circle packing determines a metric.}
Table \ref{tab:structures} summarises the relations among various structures on a mesh.

\begin{table}[ht]
    \footnotesize
    \caption{Relation among structure variables used in our algorithm}
    \hspace{-20mm}
    \begin{tabular}{|c|cccccc|}
    \hline
         &  mesh & metric & curvature & circle packing & conformal str. & embedding \\
         & (panel connection) & (panel shape) & (angle defect) &  &  & (panel location) \\
         \hline
        symbol &  $\mathcal{V},\mathcal{E},\mathcal{F}$ & $\{l_{ij}\mid e_{ij}\in \mathcal{E}\}$
        & $\{K_i\mid i\in \mathcal{V}\}$ &
        $\{r_i\mid i\in \mathcal{V}\}$ & $\{\eta_{ij}\mid e_{ij}\in \mathcal{E}\}$ & $\{v_i\in \R^3\}$\\
        \begin{tabular}{c}
        determined \\ by \end{tabular}& & \multirow{2}{*}{\begin{tabular}[c]{@{}l@{}}$\{v_i\}$, or\\ 
        $\{\eta_{ij}\}$ and $\{r_i\}$\end{tabular}}  & $\{l_{ij}\}$ &
        \multirow{2}{*}{\begin{tabular}[c]{@{}l@{}}$\{l_{ij}\}$ \\ (non canonical) \end{tabular}}  
        & $\{l_{ij}\}$ and $\{r_i\}$ &  \\
        & & & & & & \\
        \hline
    \end{tabular}
    \label{tab:structures}
\end{table}

When we would like to control a certain property, say the Gaussian curvature, 
we can deform upstream properties such as edge lengths and circle packing radii.
This allows us flexibility in choosing design variables when the problem is formulated in terms of an optimisation problem.

\begin{figure*}[ht]
	\centering
	\includegraphics[width=0.99\textwidth]{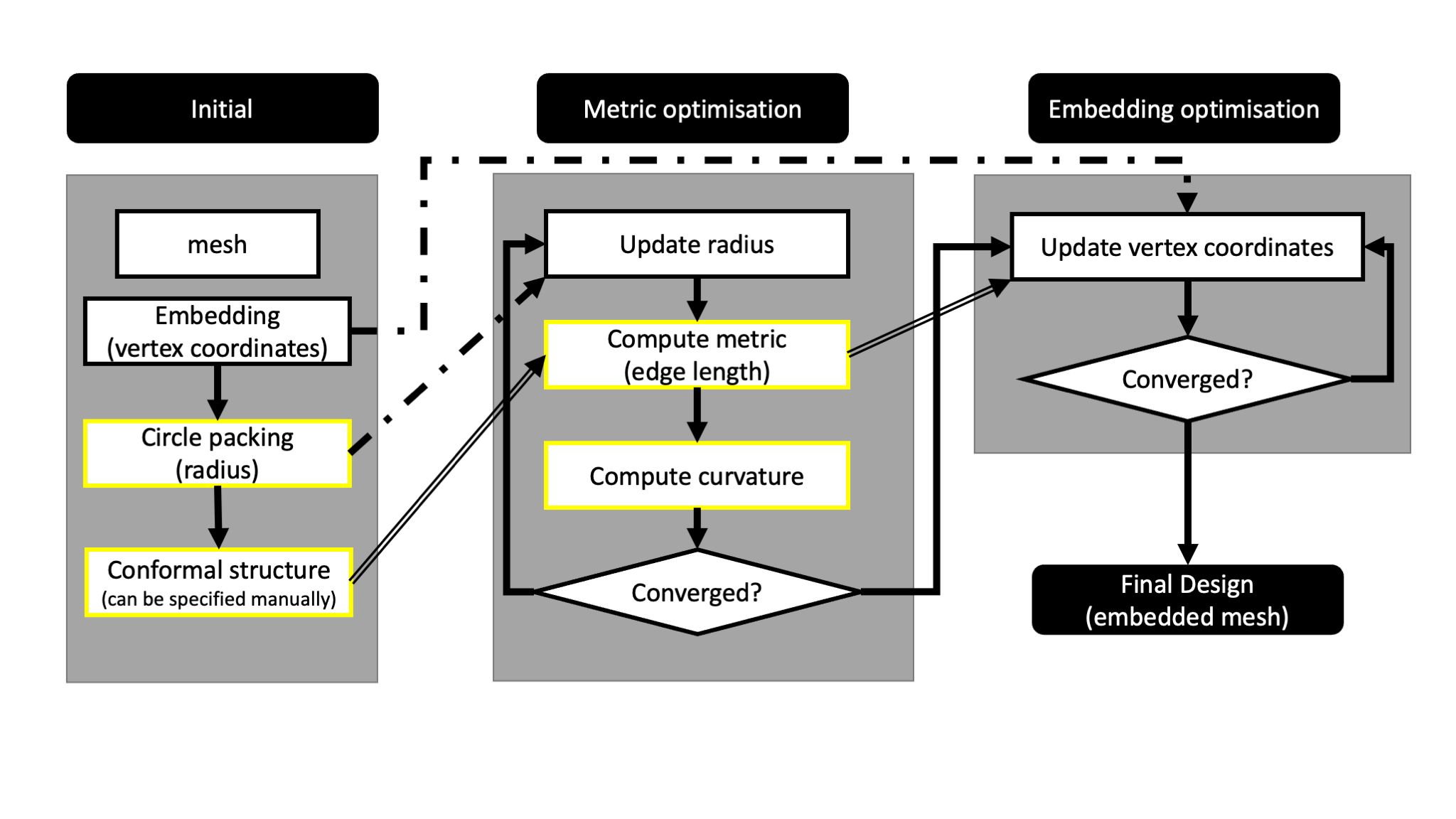}
	\caption{Overview of the proposed method.
	The dashed arrows indicate the values are used as the initial solution.
	The double-line arrows indicate the values are used but not changed.}
\label{Fig:algorithm}
\end{figure*}

Our proposed method consists of two steps; metric optimisation and embedding optimisation
(see \reffig{Fig:algorithm}).
Given a mesh and a conformal structure, 
the metric optimisation finds a metric $\{l_{ij}\}$ having the user-specified Gaussian curvatures 
and the user-specified lengths on a subset $\mathcal{E}_{\text{fix}}$ of edges (typically, the set of the boundary edges). {This part determines the approximate shape of the panels that are used to construct the final surface.}
Then, the embedding optimisation finds an embedding $\{v_i\in \R^3\}$ having the metric $\{l_{ij}\}$
obtained in the previous step and the user-specified coordinates on 
a subset $\mathcal{V}_{\text{fix}}$ of vertices (typically, the set of the boundary vertices).
{This part determines the actual position of each panel.}
The final design is obtained as the embedded mesh in the Euclidean $3$-space $\R^3$.

\subsection{Metric optimisation}

Assume that a mesh $(\mathcal{V}, \mathcal{E}, \mathcal{F})$
together with user-specified target Gaussian curvatures $\{\bar{K}_i\mid i \in \mathcal{V}_K\}$
on a subset $\mathcal{V}_K$ of vertices (typically, the set of interior vertices) is given.
Moreover, the user specifies a conformal structure $\{\eta_{ij}\}$.
For example, if the mesh has a uniform vertex valence of six, setting 
the constant value $\eta_{ij}=1$ for all edges $e_{ij}\in\mathcal{E}$
encourages the final mesh to have a regular tessellation with equilateral triangular panels.
Alternatively, 
the conformal structure $\eta_{ij}$ can be computed from the metric 
of initial design by \eqref{Eq:inversive} and \eqref{Eq:eta}.
In this case, the final mesh will be in the same conformal class as the initial one,
and hence, the shape of panels are expected to be preserved.
Moreover, 
the user can specify certain edges $e_{ij}\in \mathcal{E}_{\textrm{fix}}$
to have the specified lengths $\bar{l}_{ij}$.
For example, if we would like the positions of the boundary vertices
fixed to $\{\bar{v}_i \mid i \in \mathcal{V}_\text{bd}\}$ in the final embedding,
we set $\mathcal{E}_{\text{fix}}=\mathcal{E}_{\text{bd}}$
and $\bar{l}_{ij}=|\bar{v}_i-\bar{v}_j|$ at this stage.

To achieve these goals, we solve the following non-linear least-squares problem:
\begin{align}
	\label{Eq:opt_metric}
	{\rm Find} & \quad u_i, \quad (i\in \mathcal{V}) \nonumber\\
	\textrm{Minimize} & \quad
	\sum_{i\in \mathcal{V}_{K}} (K_i(\mathbf{u})-\bar{K}_i)^2 
	+ \lambda_e\sum_{e_{ij}\in
	\mathcal{E}_{\textrm{fix}}}(l_{ij}(\mathbf{u})^2-\bar{l}_{ij}^2)^2,
	\end{align}
where the first term is the modified Ricci energy \eqref{Eq:modifiedEnergy},
and $l_{ij}(\mathbf{u})$ is computed by \eqref{Eq:circle1} and $r_i=\exp{u_i}$.
The hyper-parameter $\lambda_e\ge 0$ controls the trade-off between 
conformity to the specified Gaussian curvature and the 
specified edge lengths.
Once $u_i$ are obtained as the solution to \eqref{Eq:opt_metric},
the metric $\{l_{ij}\}$ is computed by \eqref{Eq:circle1} with $r_i=\exp(u_i)$.
\begin{remark}
Instead of $u_i$, we can directly optimise the radii $r_i$
in \eqref{Eq:opt_metric}.
The variable $u_i$ is often called the conformal factor in analogy to the smooth setting.
Algorithmically, setting $r_i=\exp(u_i)$ serves as
a change of variables to ensure the positivity constraint $r_i>0$.
We may also directly optimise the edge lengths $l_{ij}$ since $K$ is a function of $l_{ij}$ as well.
However, we cannot specify the conformal class of the resulting metric. 
Moreover, ensuring the triangular inequality
can be problematic.
Our codes implement all these choices; $u_i$, $r_i$, and $l_{ij}$.
\end{remark}

\subsection{Embedding optimisation}
Now, we would like to find an embedding $\{v_i\in \R^3\}$ of the mesh
which realises the metric $\{l_{ij}\}$ obtained in the previous step
while a subset $\mathcal{V}_{\text{fix}}$ of vertices (typically, the boundary vertices) have the user-specified coordinates $\bar{v}_i$.
The problem is similar to the classical multi-dimensional scaling (MDS)~\cite{Mead1992} in which an embedding of finite points with a specified distance matrix is sought for.
Our embedding problem can be thought of as a generalisation of MDS to graphs 
where only a subset of pairs have specified distances. 
{For a triangular mesh,
there is a sophisticated method proposed in \cite{Chern},
where an isometric immersion of a triangular mesh with prescribed edge lengths is obtained.
This method optimises the orientation of each triangle represented by a unit quaternion,
and hence, it is difficult to incorporate constraints defined with vertex positions such as the boundary condition.
Since the boundary constraints is important for our purpose, 
we rely on a straightforward 
algorithm which takes the vertex position directly as its optimisation variable.
More precisely, we formulate it as a type of \emph{graph embedding} problem~\cite{cai2018}.
}
What is peculiar to our case is the fact that a uniform scaling will not change the Gaussian curvatures at the vertices.
Hence, we introduce a scaling factor $\beta>0$ and solve the following 
non-linear least-squares problem\footnote{When $\mathcal{V}_{\text{fix}}=\emptyset$, 
we set $\beta=1$ and do not optimise $\beta$ to avoid the degenerate solution of $\beta=0$ and $v_i=v_j$ for all $i,j\in\mathcal{V}$}:
\begin{align}
	\label{Eq:opt1}
	{\rm Find} & \quad \beta, v_i, \quad (i\in \mathcal{V}) \nonumber\\
	\textrm{Minimize} & \quad \sum_{e_{ij}\in \mathcal{E}}(|v_i-v_j|^2-\beta l_{ij}^2)^2
	+ \lambda_v \sum_{i\in \mathcal{V}_{\text{fix}}} |v_i-\bar{v}_i|^2,
\end{align}
where $\lambda_v\ge 0$ is a hyper-parameter controlling the trade-off between 
conformity of the Gaussian curvature and the boundary coordinates.
Other constraints can be easily incorporated by adding penalty terms to \eqref{Eq:opt1}.
For example, to enforce the convexity at internal vertices, we add
the following term with a hyper-parameter $\lambda_c\ge 0$
\begin{align}
	\label{Eq:convex}
	&\lambda_c \sum_{i\in \mathcal{V}\setminus \mathcal{V}_\text{bd}}f(\tilde{v}_{i,z}-v_{i,z}), \nonumber\\
	&\left(f(x)=\dfrac{x}{1+\exp(-x)}\right),
\end{align}
where $v_{i,z}$ is the $z$-coordinate of $v_i$
and $\tilde{v}_{i,z}$ is the mean of the $z$-coordinates of 
the vertices adjacent to $v_i$.
{The sigmoid-weighted linear function~\cite{swish} $f(\tilde{v}_{i,z}-v_{i,z})$ penalises 
$\tilde{v}_{i,z} > v_{i,z}$}, and its gradient is computed by
$f'(x)=f(x)+\dfrac{1-f(x)}{1+\exp(-x)}$.
Another example of a penalty term we can add to \eqref{Eq:opt1} is 
\begin{align}
	\label{Eq:close}
	\lambda_r\sum_{i\in \mathcal{V}}|v_i-\hat{v}_i|^2,
\end{align}
where $\hat{v}_i$ is the vertex coordinate of the initial design and $\lambda_r\ge 0$
is the hyper-parameter. 
This regularisation term encourages the final design stay close to the initial one.
More generally, we can combine the boundary constraint term and the regularisation term into 
a single weighted sum of the form $\sum_{i\in \mathcal{V}} w_i |v_i-\bar{v}_i|^2$,
where $\bar{v}_i$ is the target coordinate and $w_i\ge 0$ is the weight for each vertex.

The vertex positions of the initial design are used as the initial solution for 
the optimisation problem \eqref{Eq:opt1}.
Note that finding an embedding which realises the specified metric
is not convex, and there can be multiple global minima, as is seen in \reffig{Fig:house}. 
In this case, the choice of an initial solution affects the final solution.
{
This problem can be mitigated to some extent by regularising the optimisation by adding, for example, the Laplacian smoothing term to the target function \cite{nealen}.
}

There exists a metric for a mesh with a conformal structure 
which realises the specified Gaussian curvature 
if and only if \eqref{Eq:gauss-bonnet} and \eqref{Eq:admissible} are satisfied. 
Furthermore, when such a metric exists, it is unique up to a uniform scaling.
However, there may not exist an embedding which realises the metric~\cite{Erdos80}.
Also, there can be multiple non-congruent embeddings which realise the metric.
Therefore, the metric optimisation is theoretically more troublesome, and 
this part is separated from the whole procedure in
our scheme.

\section{Numerical examples}\label{sec:example}
\begin{table*}[thb]
    \centering
    \caption{Error analysis and computation time of examples}
    \begin{tabular}{c|ccccc}
        \hline
         & $A_K$ (radian)& $A_v$ (meter)& $A_\theta$ (degrees)& metric opt. (sec) & embedding opt. (sec) \\
        \hline
        Ex1a  & $1.37\times 10^{-4}$ & $8.09\times 10^{-2}$ & 0.49 & 0.28 & 0.32 \\
        Ex1b  & $6.07\times 10^{-5}$ & $1.96\times 10^{-2}$ & 0.99 & 0.28 & 0.55 \\
        Ex1c  & $9.12\times 10^{-6}$ & $6.80\times 10^{-3}$ & 0.13 & 0.60 & 0.26 \\
        Ex1d  & $1.72\times 10^{-2}$ & $6.52\times 10^{-4}$ & 0.55 & 0.33 & 0.31 \\
        Ex2   & $3.48\times 10^{-4}$ & $8.18\times 10^{-2}$ & 0.50 & 0.14 & 0.59 \\
        Ex3  & $4.24\times 10^{-4}$ & $8.31\times 10^{-2}$ &  0.54 & 0.24 & 1.87 \\
        Ex4a & $2.06\times 10^{-6}$ & $1.19\times 10^{-3}$ & 0.06 & 0.53 & 4.08 \\
        Ex4b & $4.97\times 10^{-6}$ & $2.89\times 10^{-3}$ & 0.08 & 0.53 & 8.32 \\
         \hline
    \end{tabular}
    \label{Tab:Ex1a_error}
\end{table*}

In this section, we present several numerical examples to validate the proposed method.
{
We are mainly interested in dome-shaped surfaces 
with a high structural performance covering a large column-free space.
To design a high structural performance covering surface built on a specified boundary, 
minute control over the Gaussian curvature, panel shape, and the boundary positions is important.}
We evaluate the results in terms of the mean absolute errors of
the Gaussian curvatures, the coordinates of the boundary vertices, and the angles of the faces
of the final design against the target values:
{\footnotesize
\begin{align}
    \label{Eq:performace2}
    A_{K} &= \frac{1}{|\mathcal{V}_K|}\sum_{i\in \mathcal{V}_K}{ |K_{i} - {\bar K}_{i}|  }, \nonumber \\
    A_{v} &= \frac{1}{|\mathcal{V}_{\text{fix}}|}\sum_{i\in \mathcal{V}_{\text{fix}} }{ |v_i-\bar{v}_i| }, \nonumber \\
    A_\theta &= \frac{1}{3|\mathcal{F}|}\sum_{f_{ijk} \in \mathcal{F}}\left( |\theta_i^{jk} - \bar{\theta}_i^{jk}| + |\theta_j^{ki} - \bar{\theta}_j^{ki}| 
    +|\theta_k^{ij} - \bar{\theta}_k^{ih}| \right),
\end{align}}
where ${\bar\theta}_i^{jk}$ are the target corner angles of the triangular panels,
and $\theta_i^{jk}$ are those of the final mesh.
When a conformal structure is given by an initial design, ${\bar\theta}_i^{jk}$ are 
the corner angles of the initial mesh.
When a conformal structure is specified manually, ${\bar\theta}_i^{jk}$ depends on the mesh topology.

All the experiments are performed by our Python script\footnote{\url{https://github.com/shizuo-kaji/ricci_flow}} using SciPy optimisers~\cite{Virtanen2020}.
There is a trade-off among above-mentioned evaluation indicators. 
We can control the trade-off by modifying the hyper-parameters 
such as $\lambda_e$, $\lambda_v$, and stopping criteria for optimisation.
We terminate the optimisation steps when
the norm of the gradient gets less than $10^{-6}$.
Timing is measured on a MacBook Pro with a M1 core using a single core.
The results are summarised in Table \ref{Tab:Ex1a_error}.

\subsection*{Example 1: CGC surface with a hexagonal plan}
\begin{figure*}[ht]
	\centering
		\includegraphics[width=0.25\textwidth]{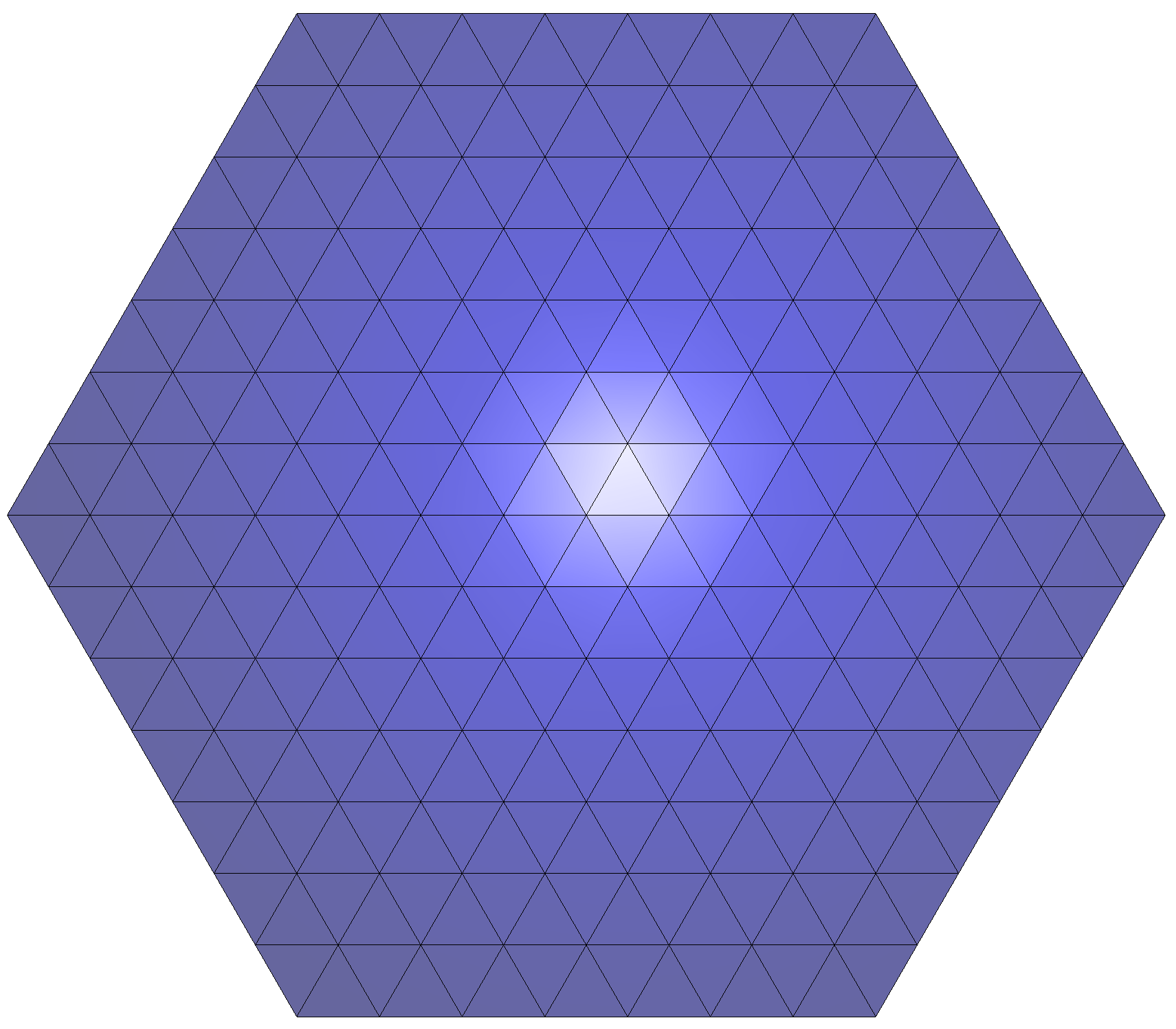}
		\includegraphics[width=0.35\textwidth]{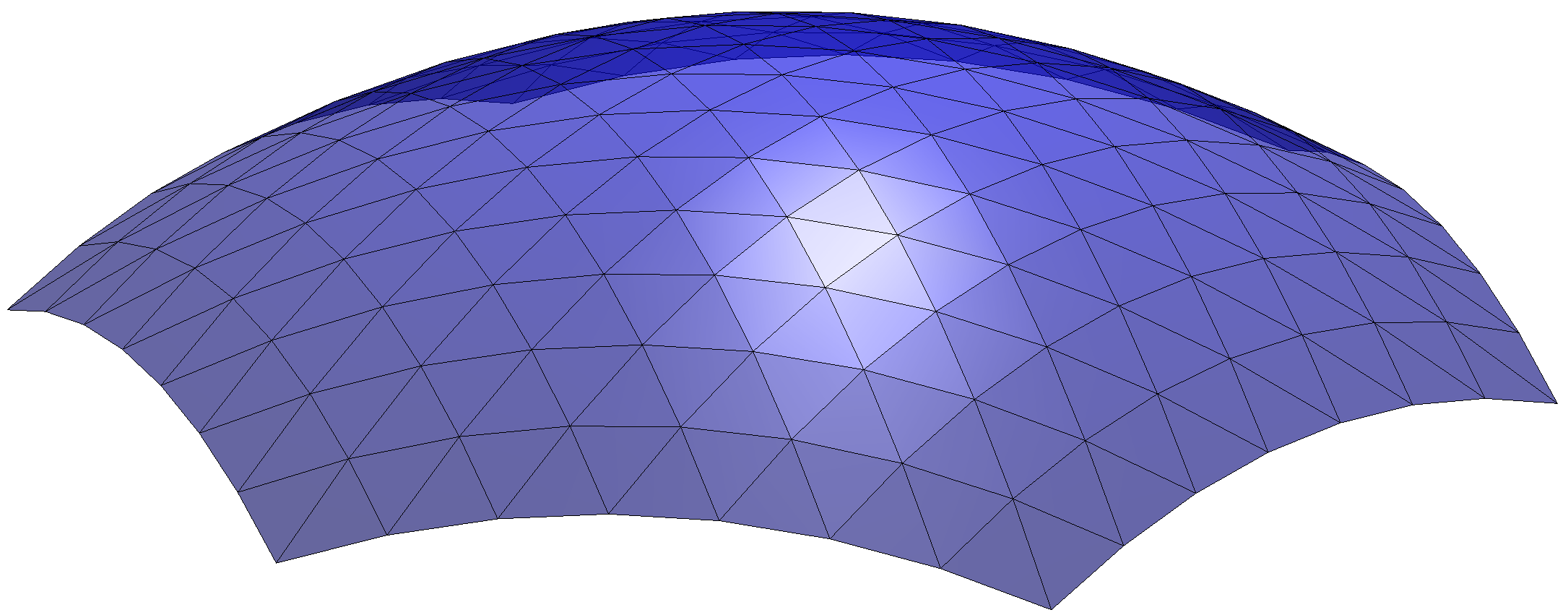}
		\includegraphics[width=0.33\textwidth]{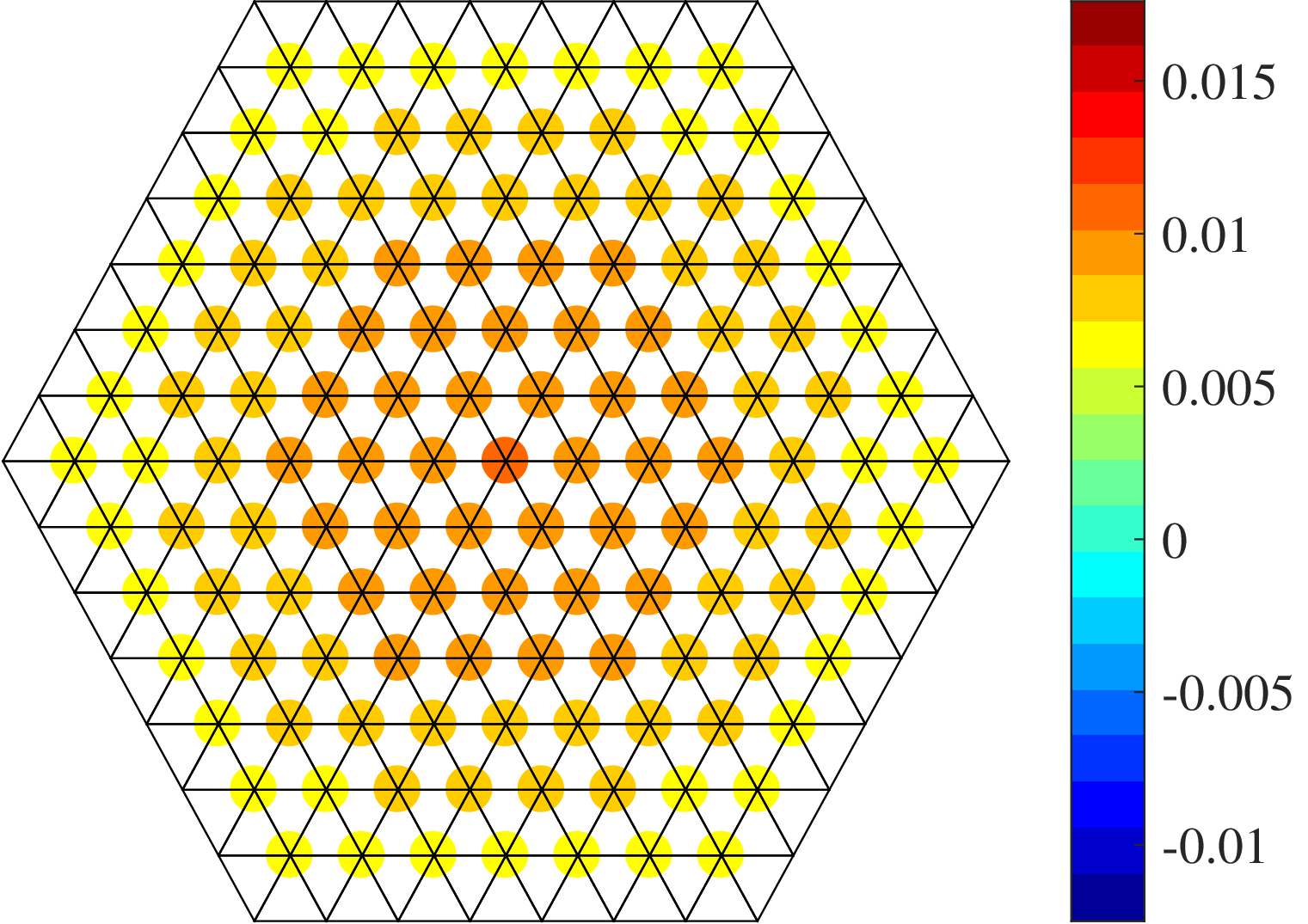}
	\caption{Initial settings of Example 1.
	From left to right: top view, perspective view, Gaussian curvatures.}
	\label{Fig:Ex1}
\end{figure*}

We begin with a simple example of a constant Gaussian curvature (CGC) surface with a hexagonal plan.
The initial geometry is shown in \reffig{Fig:Ex1}.
Its span is 30.0m and its height is 10.0m.
This surface is composed of 169 vertices (of which 42 are boundary vertices), 462 edges, and 294 faces.

For comparison, we performed several experiments with the same mesh but
varying one design target at a time.
\begin{itemize}
    \item 
For Ex1a, 
we set $\eta_{ij}=1$ aiming at equilateral panels (that is, ${\bar\theta}_i^{jk}=\pi/3$).
The total target Gaussian curvature for interior vertices is set to 1.5,
and hence, the target Gaussian curvature ${\bar K}_i$ of each interior vertex is 0.011811.
The target positions of the boundary vertices are set to the same as in the initial geometry with $\lambda_v=\lambda_e=0.01$.

\item In Ex1b,
the total target Gaussian curvature for interior vertices is set to 3.0,
and hence, the target Gaussian curvature ${\bar K}_i$ of each interior vertex is 0.023622. Other conditions are the same as Ex1a.

\item In Ex1c, we calculate $\eta_{ij}$ from the initial geometry by \eqref{Eq:inversive} and \eqref{Eq:eta} so that the final mesh has
similar triangle panels as the initial ones.
Other conditions are the same as Ex1a.

\item In Ex1d, we set $\lambda_v=\lambda_e=100$ to stress the position condition of the boundary vertices.
Other conditions are the same as Ex1a.
\end{itemize}

\begin{figure*}[ht]
	\hspace{-20mm}
	\begin{tabular}{cccc}
		\includegraphics[width=0.28\textwidth]{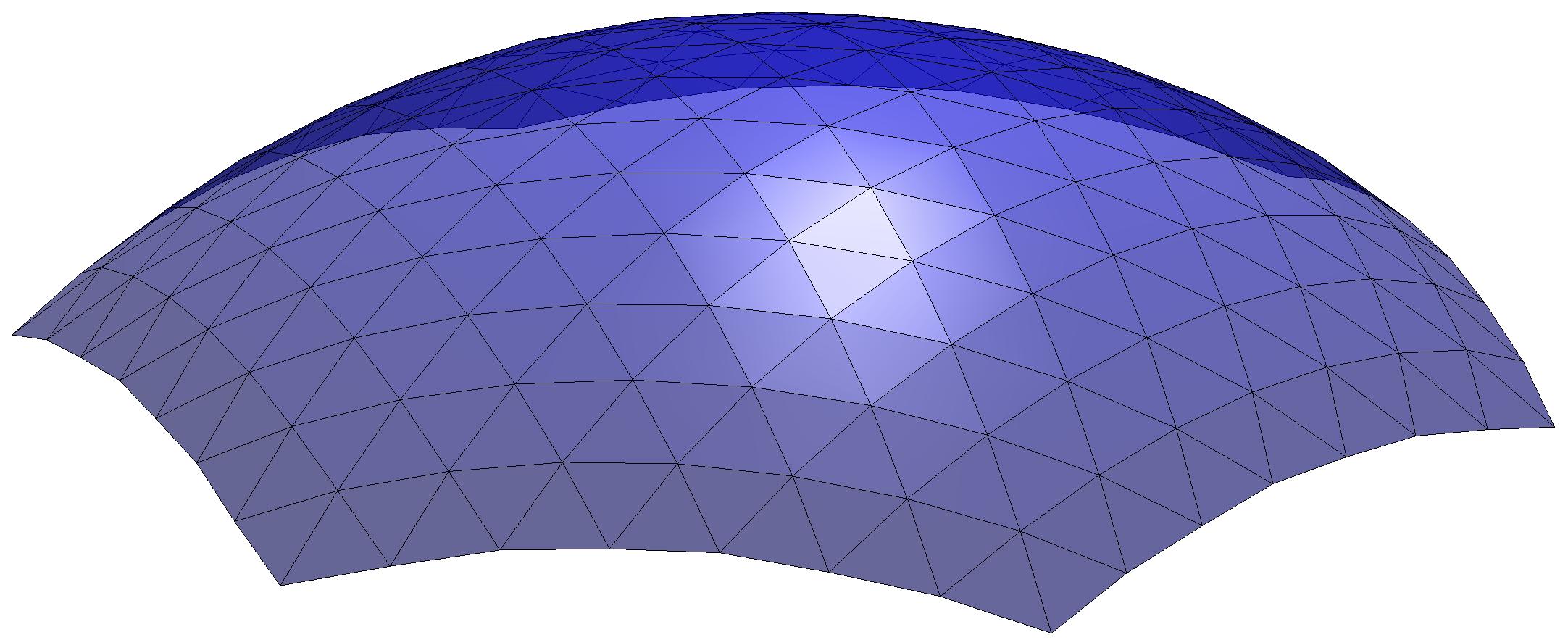} &
		\includegraphics[width=0.28\textwidth]{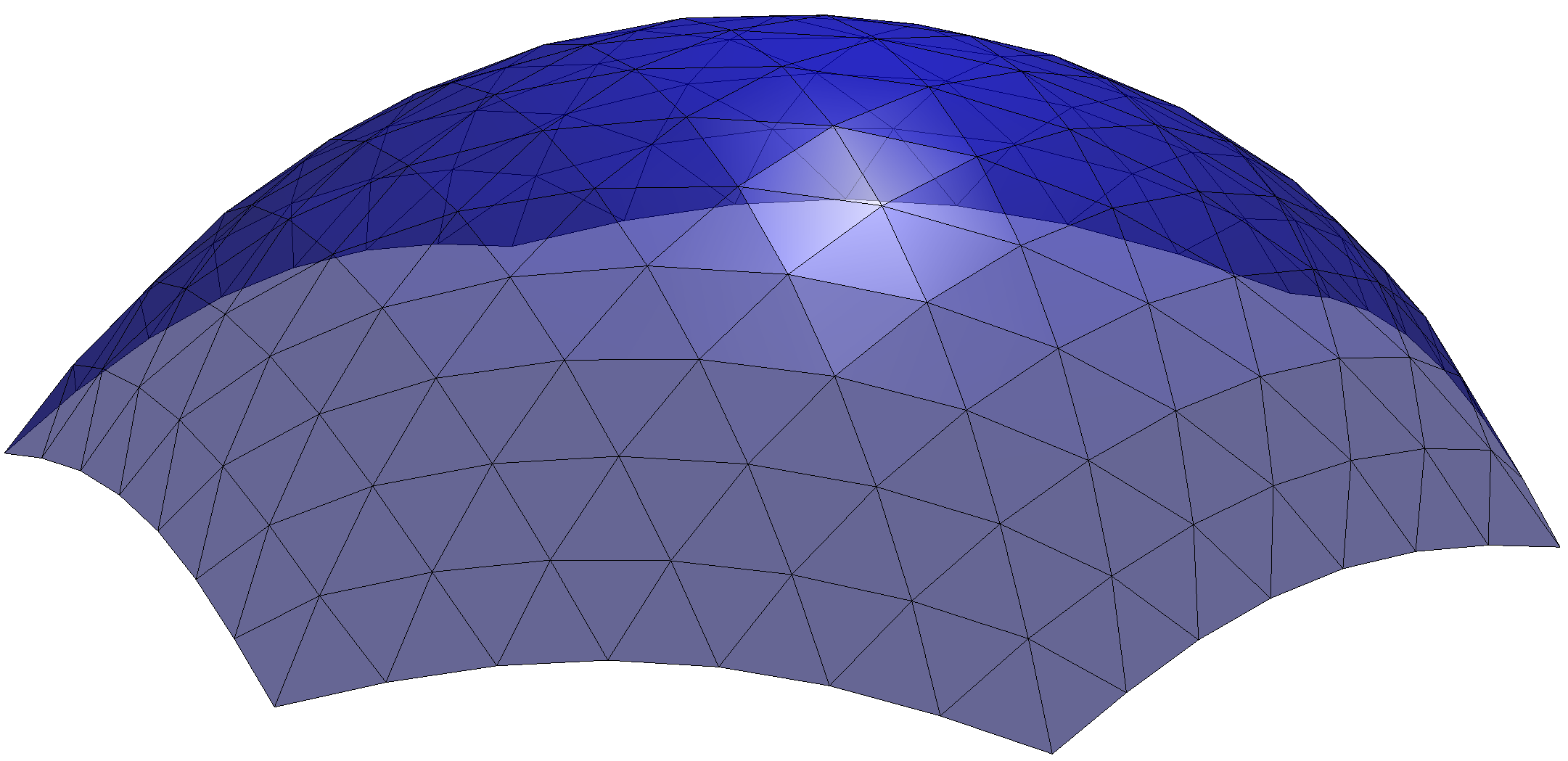} &
		\includegraphics[width=0.28\textwidth]{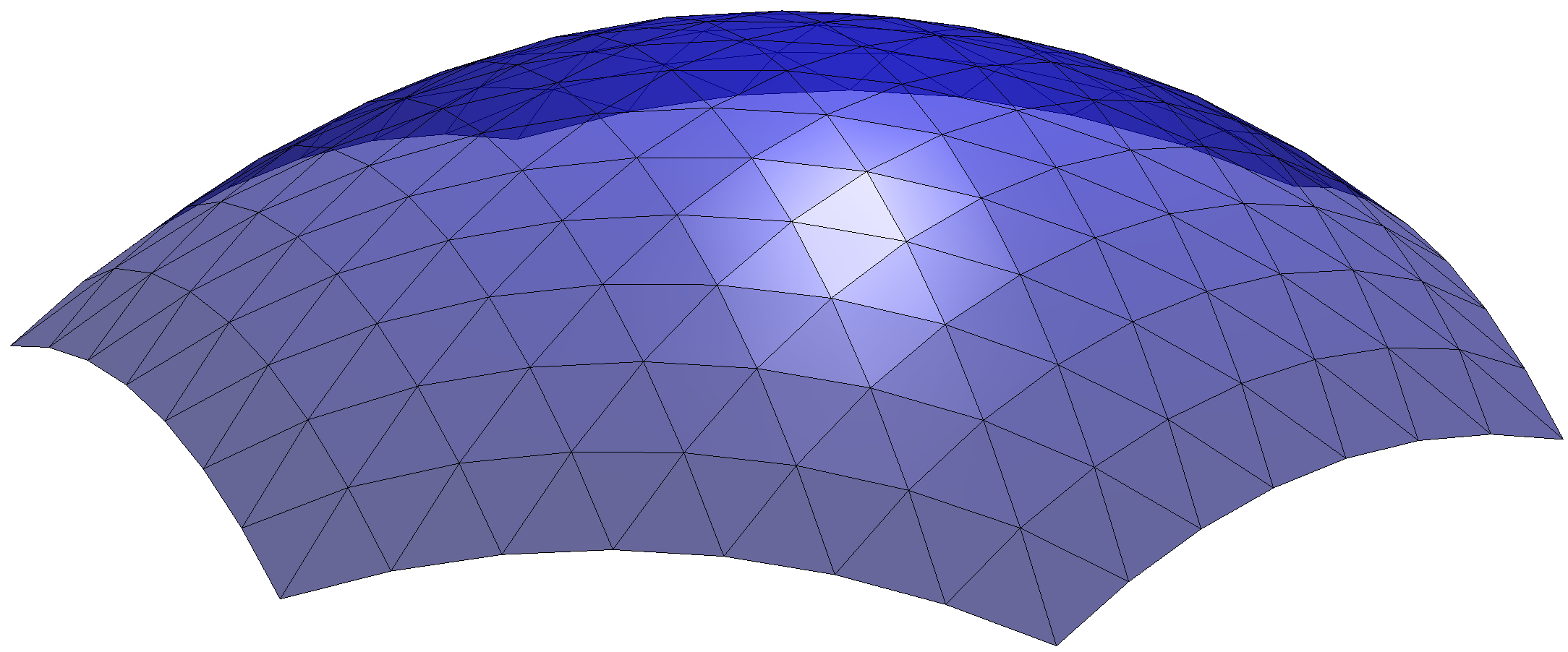} &
		\includegraphics[width=0.28\textwidth]{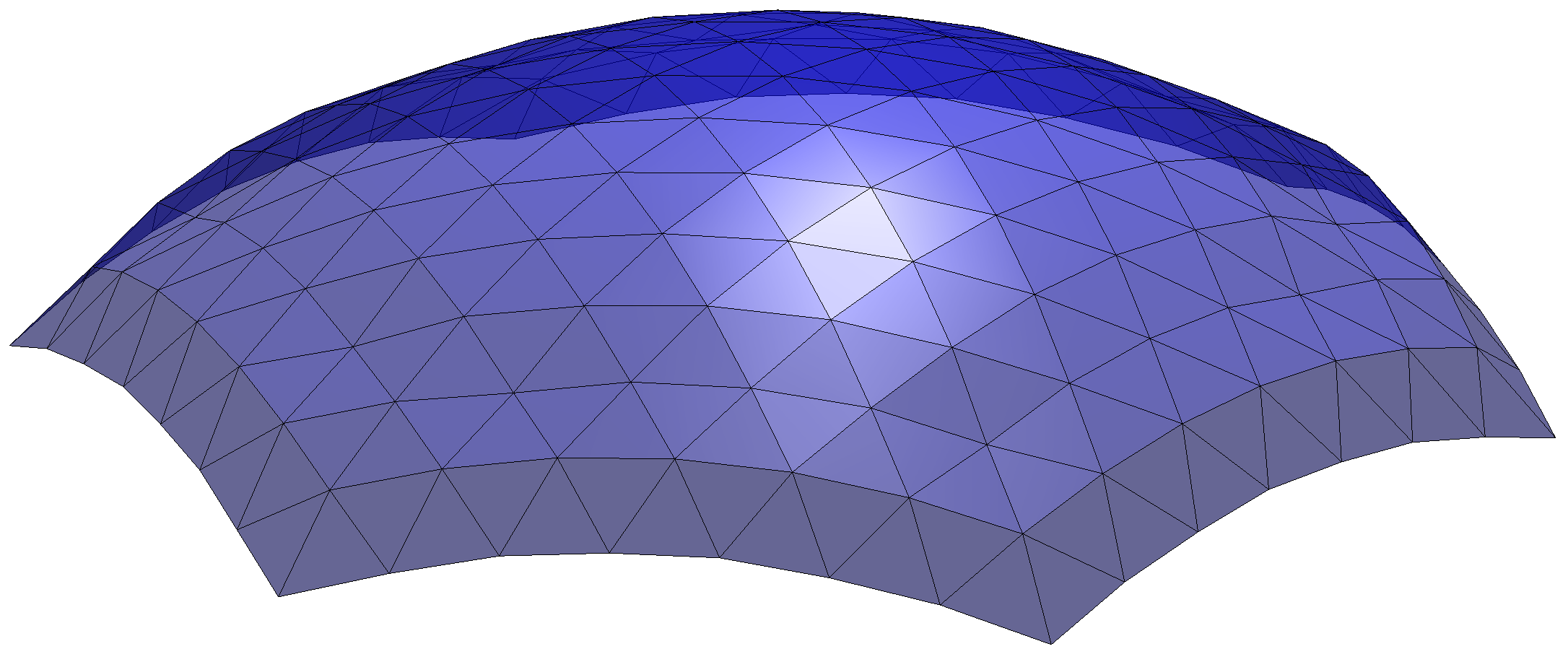} \\
		 Ex1a &  Ex1b &  Ex1c & Ex1d 
	\end{tabular}
	\caption{Final geometry of Example 1 obtained by running our algorithm 
	on Fig. \ref{Fig:Ex1} 
	with different constant target Gaussian curvatures, target conformal structures, 
	and the weights for the boundary constraints.}
	\label{Fig:Ex1_geometry}
\end{figure*}

\begin{figure*}[ht]
	\centering
	\hspace{-20mm}
		\includegraphics[width=0.55\textwidth]{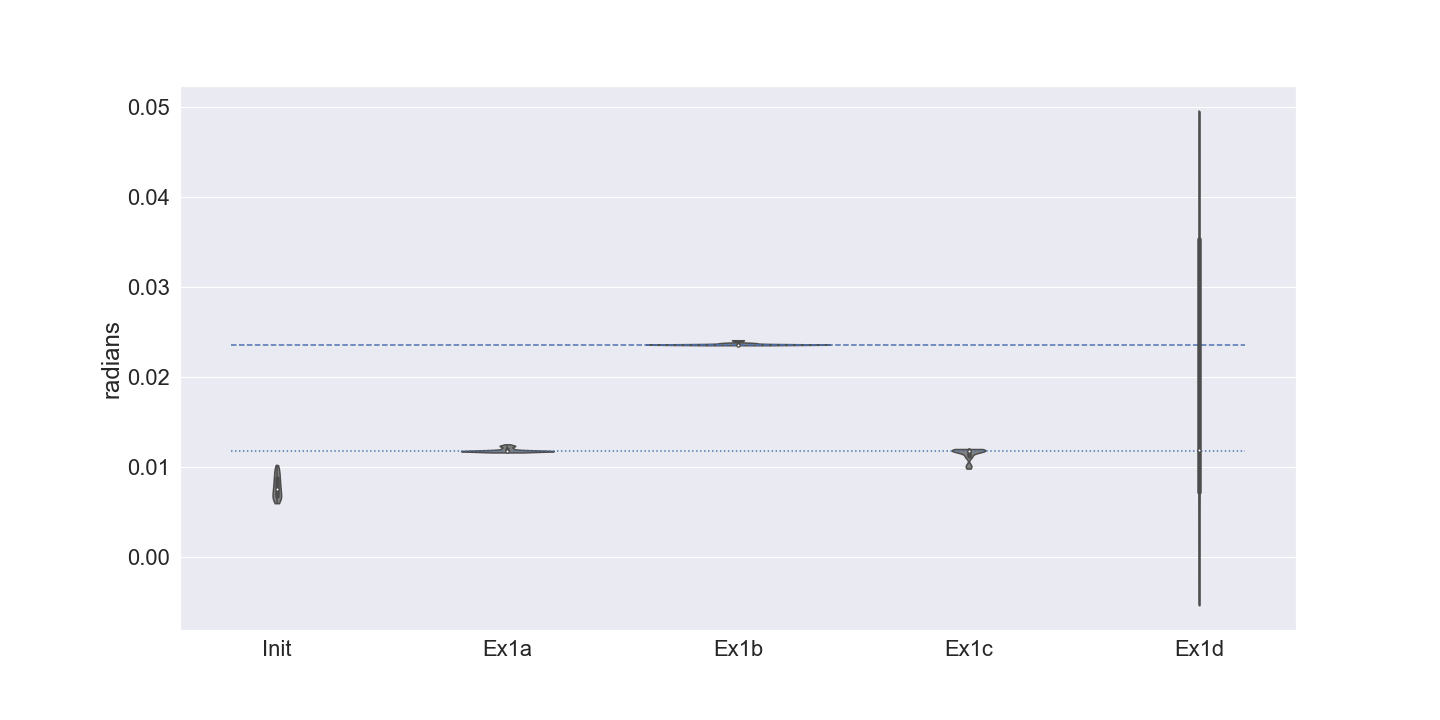}
		\includegraphics[width=0.55\textwidth]{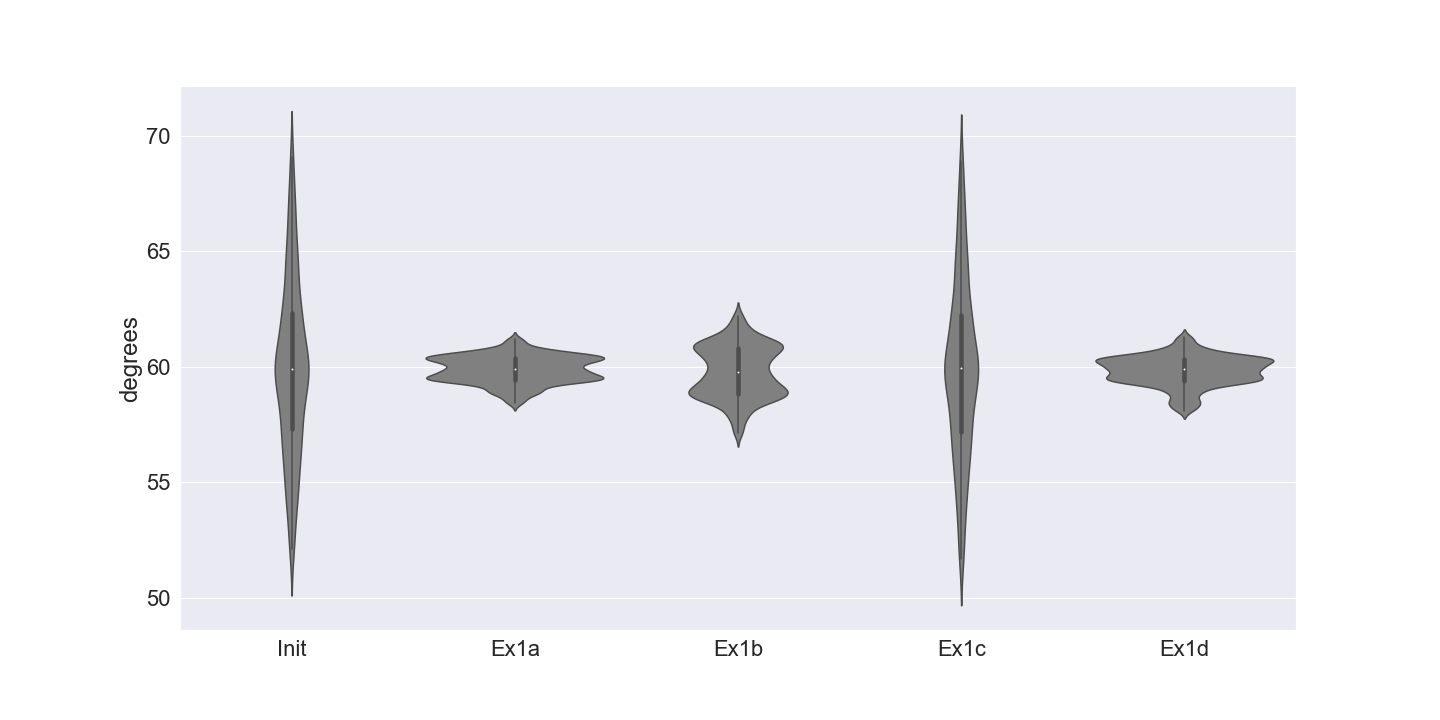}
	\caption{(Left) Distribution of Gaussian curvature of Example 1.
	 Target curvatures are set to ${\bar K}_i=0.011811$ (dotted line) for Ex1a, Ex1c, and Ex1d,
	 and ${\bar K}_i=0.023622$ (dashed line) for Ex1b.
	 (Right) Distribution of angles of Example 1.
	Target angles are set to 60 degrees for Ex1a, Ex1b, and Ex1d
	while they are set to the ones in the initial mesh in Ex1c.}
	\label{Fig:Ex1_curvature}
\end{figure*}

The final geometries are shown in \reffig{Fig:Ex1_geometry}.
The distributions of the Gaussian curvatures and the corner angles at interior vertices for the initial and final meshes
are shown in the violin plots~\cite{Hintze1998} in \reffig{Fig:Ex1_curvature}.

In Ex1a, Ex1b, and Ex1c, the distributions of the Gaussian curvatures
concentrate around the target values.
In Ex1d, the stringent constraint increased the precision of the boundary location
but this is achieved at a large cost of the precision of the Gaussian curvatures.
This suggests that methods which only allow the user to specify the exact boundary locations will fail.
Our method offers control over the trade-off through the choice a hyper-parameter.

In the initial geometry, the corner angles of the panels range from 50 to 70 degrees,
while in Ex1a, Ex1b, and Ex1d, they are corrected to 60 degrees within 3 degrees of margins,
resulting in a near regular tessellation in the final geometry.
On the other hand, in Ex1c, the distribution of the corner angles is close to that of the initial geometry as desired.

\begin{figure*}[ht]
	\centering
		\includegraphics[width=0.8\textwidth]{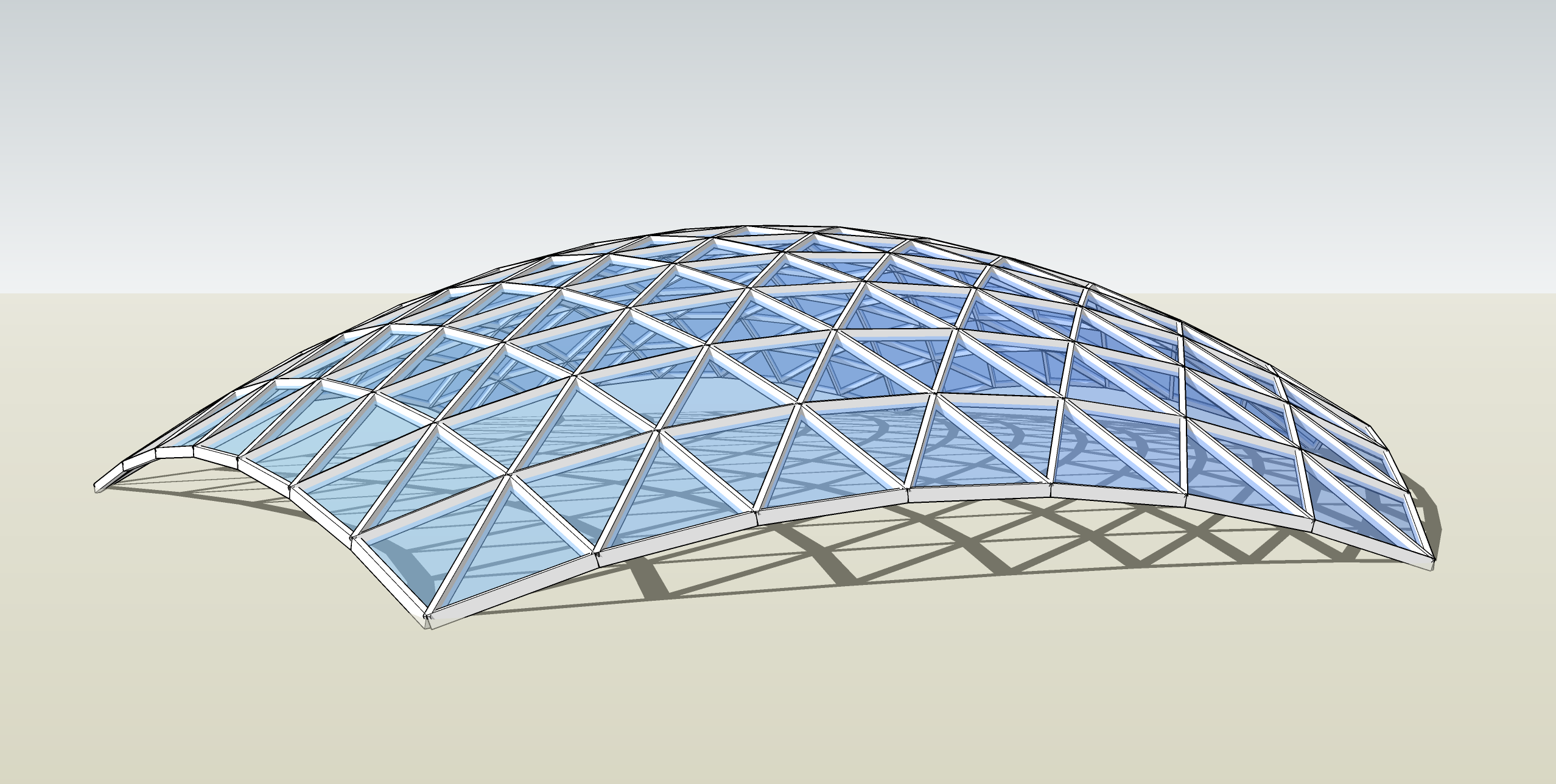}
	\caption{Rendering of the final design Ex1a.}
	\label{Fig:Ex1a_rendered}
\end{figure*}

\subsection*{Example 2: surface with irregular mesh}
\begin{figure*}[ht]
	\centering
	\begin{tabular}{ccc}
		\includegraphics[width=0.25\textwidth]{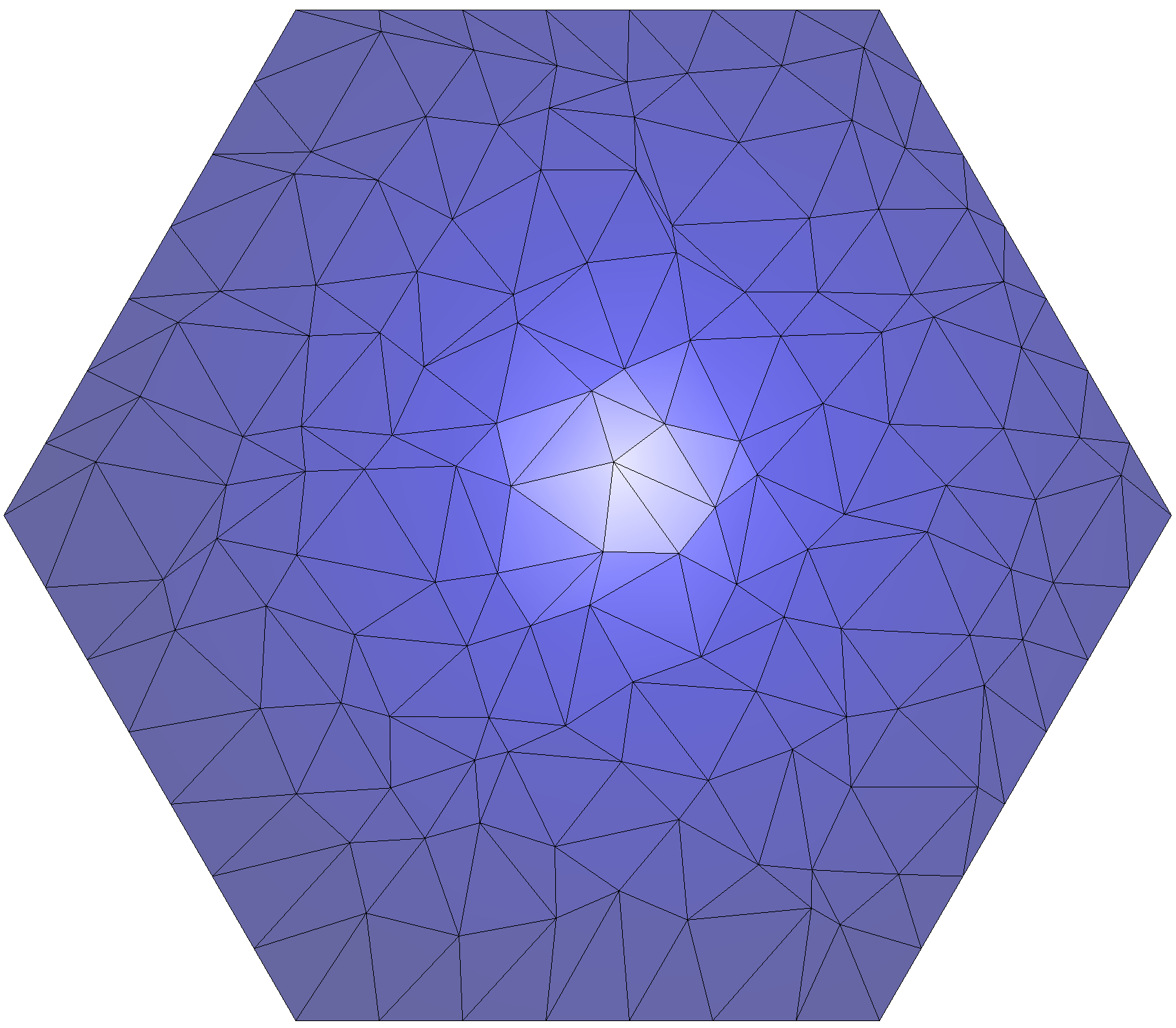} &
		\includegraphics[width=0.35\textwidth]{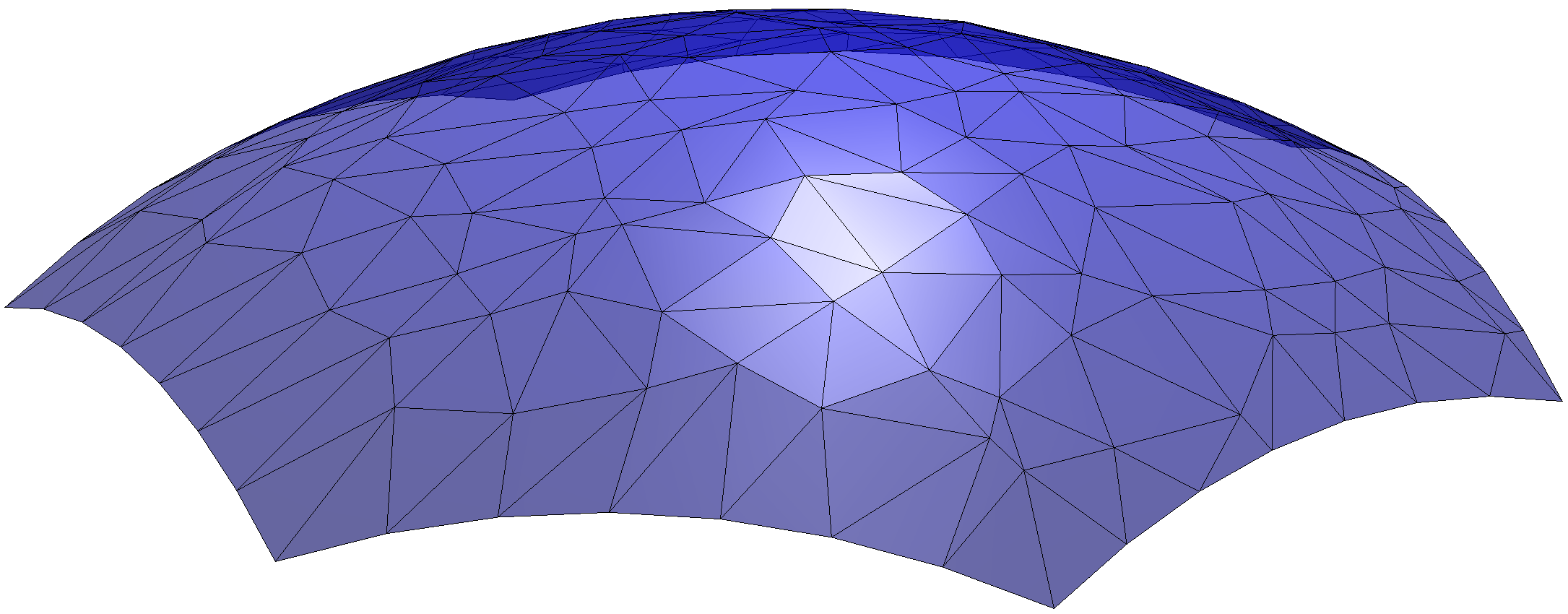} &
		\includegraphics[width=0.3\textwidth]{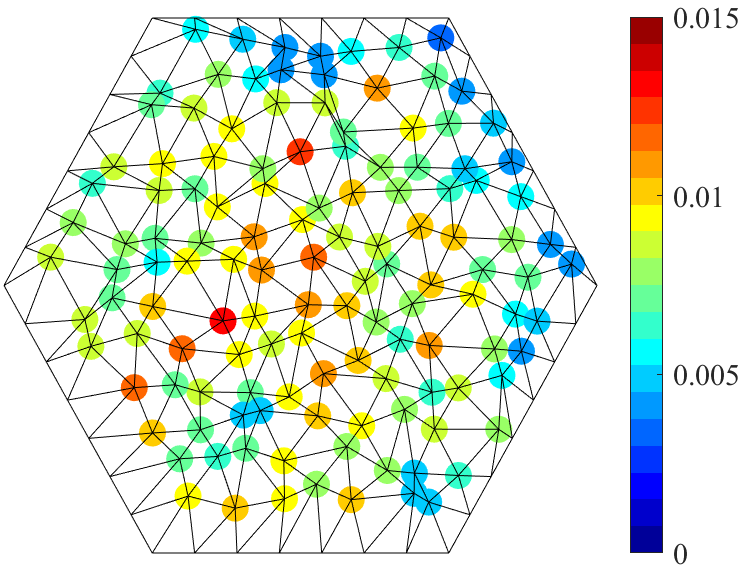} \\
	\end{tabular}
	\caption{Initial settings of Example 2, which comprise irregular triangles.
		From left to right: top view, perspective view, Gaussian curvatures.}
	\label{Fig:Ex2_initial}
\end{figure*}

\begin{figure*}[ht]
	\centering
	\begin{tabular}{cccc}
		\includegraphics[width=0.20\textwidth]{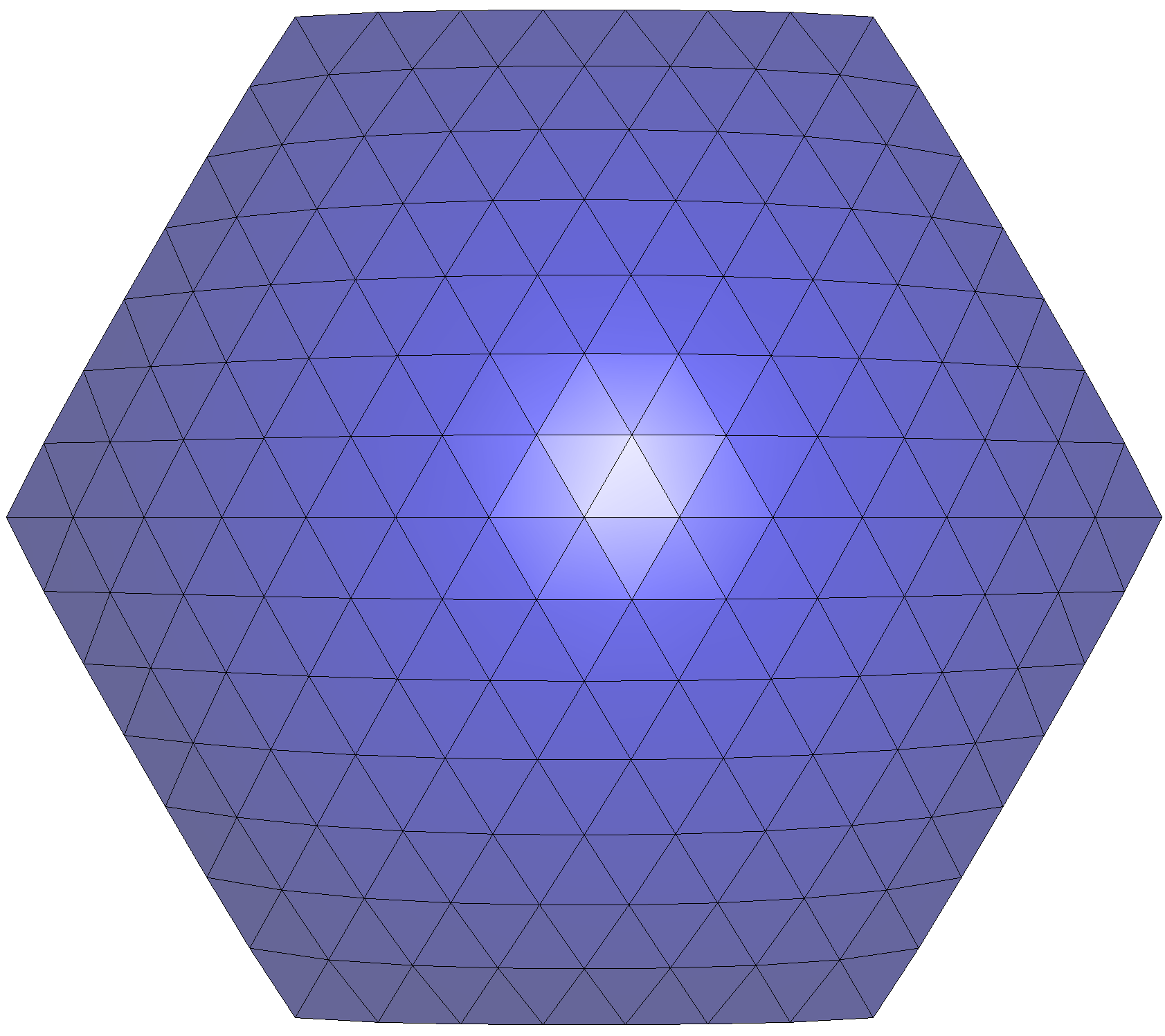} &
		\includegraphics[width=0.30\textwidth]{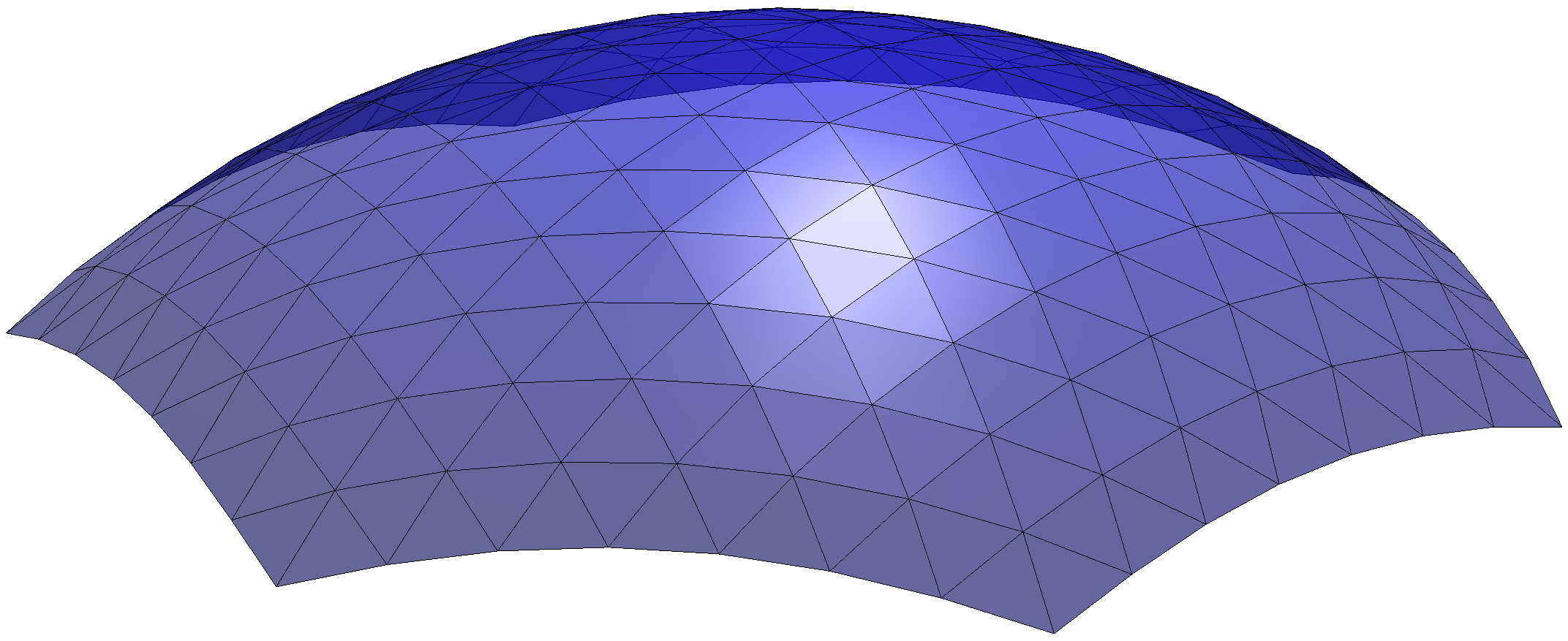} &
		\includegraphics[width=0.25\textwidth]{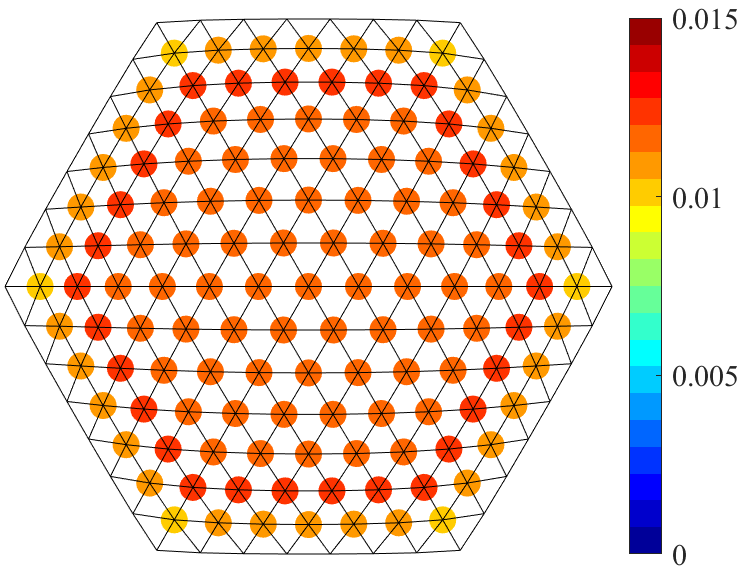} &
		\includegraphics[width=0.25\textwidth]{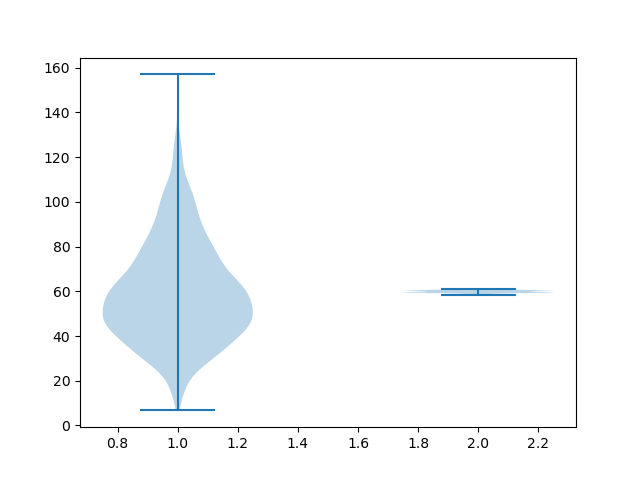} \\
	\end{tabular}
	\caption{Final results of Example 2 
	 obtained by running our algorithm on Fig. \ref{Fig:Ex2_initial}.
		From left to right: top view, perspective view, Gaussian curvatures,
		and distribution of angles of the initial and final geometries.
		The final geometry has almost constant Gaussian curvatures and 
		a nearly regular tessellation.}
	\label{Fig:Ex2_final}
\end{figure*}

To further illustrate the capacity of the proposed method in transforming an irregular initial mesh
into a regular one, 
we consider the initial geometry as shown in \reffig{Fig:Ex2_initial},
where the $xy$-coordinates of the interior vertices are randomly perturbed from those specified in Example 1.
The minimum and maximum Gaussian curvatures at the interior vertices of the initial geometry are respectively
0.0035 and 0.0129.
Moreover, the distribution of angles is illustrated in \reffig{Fig:Ex2_final},
ranging from about $5^{\rm o}$ to about $160^{\rm o}$.

The parameters of the algorithms are set to the same as Ex1a,
except that the coefficient for convexity $\lambda_c$ is set to 0.1.
In particular,
the target Gaussian curvature ${\bar K}_i$ of each interior vertex is 0.011811.
The final geometry is shown in \reffig{Fig:Ex2_initial}.
The minimum and maximum Gaussian curvatures at the interior vertices of the initial geometry are respectively
0.0101 and 0.0123,
which are close to their target values.
Moreover, as can also be observed from the angle distribution of the final geometry as in \reffig{Fig:Ex2_final},
the mesh becomes nearly regular with almost isosceles triangles.

\subsection*{Example 3: surface with non-uniform Gaussian curvature}
\begin{figure*}[ht]
	\centering
		\includegraphics[width=0.25\textwidth]{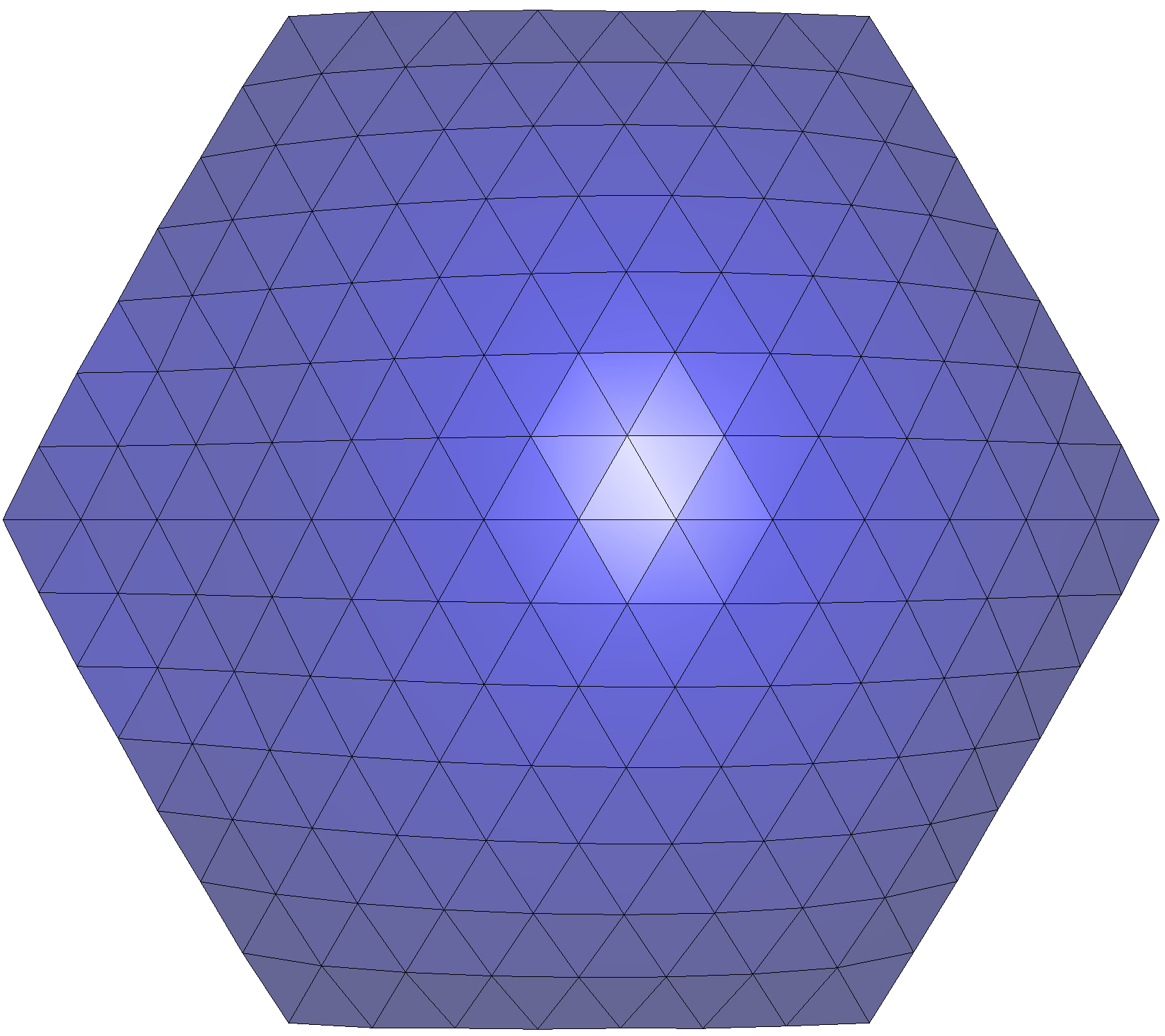} 
		\includegraphics[width=0.35\textwidth]{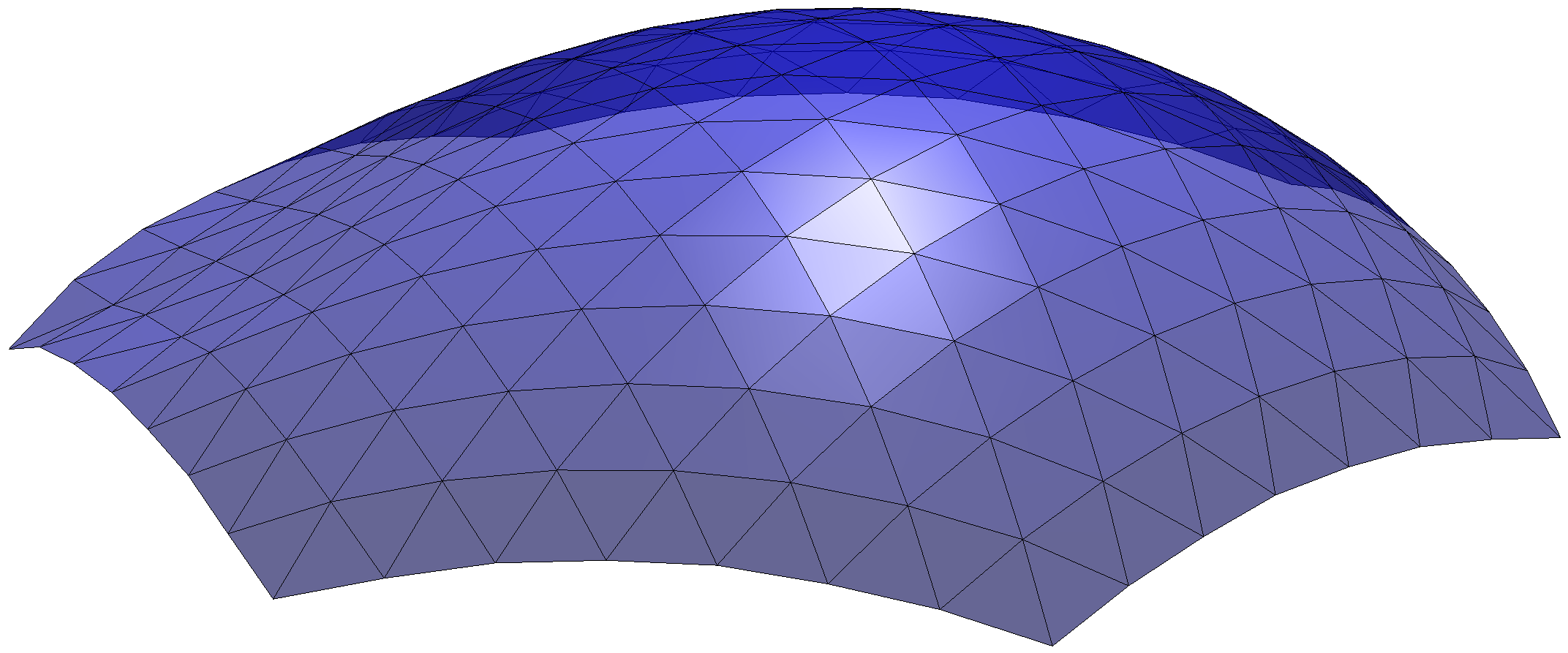} 
			\includegraphics[width=0.35\textwidth]{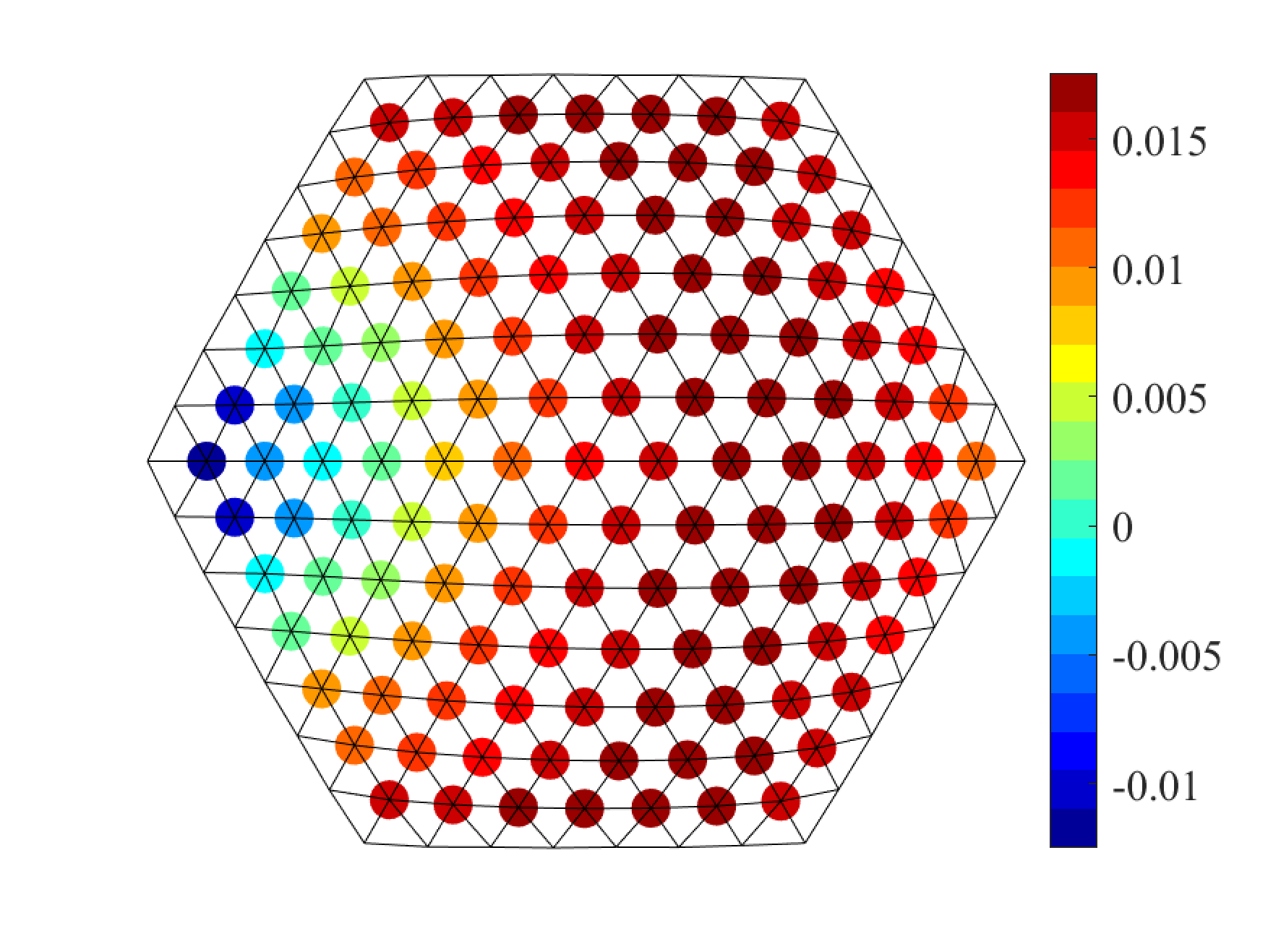} 
	\caption{Final results of Example 3, which has smoothly varying Gaussian curvatures with
	both positively and negatively curved portions.
		From left to right: top view, perspective view, Gaussian curvatures.}
	\label{Fig:Ex3_geometry}
\end{figure*}

In Example 3, we are to design a surface with varied target Gaussian curvatures 
taking positive values at some vertices and negative values at the others.
The initial geometry of Example 1 is used in this example.
We set $\eta_{ij}=1$ aiming at equilateral panels (that is, ${\bar\theta}_i^{jk}=\pi/3$).
The target Gaussian curvatures are assigned by a quadratic curve defined as follows:
\begin{align}
	{d}_i &= -\frac{\sqrt{(x_i-x_0)^2 + y_i^2}}{2x_0},\quad (0\leq d_i\leq 1) \nonumber\\
	{\bar K}_i &= -\frac{2c}{b^2}(d_i-b)^2 + c,\quad (i\in \mathcal{V}\setminus \mathcal{V}_\text{bd}),
\end{align}
where $(x_i,y_i)$ are the $(x,y)$-coordinates of the initial embedding, and
$c=0.0168$, $b=0.7$, and $x_0=-12.86$.
The target Gaussian curvature 
attains the maximum value $c$ at $d_i=b$, and the minimum value $-c$ at $d_i=0$.
The total target Gaussian curvature of the interior vertices is 1.5.
The target positions of the boundary vertices are set to the same as in the initial geometry
with $\lambda_v=\lambda_e=0.01$.
\reffig{Fig:Ex3_geometry} shows the final geometry of the surface.

\subsection*{Example 4: surface with openings}
\begin{figure*}[ht]
	\centering
		\includegraphics[width=0.4\textwidth]{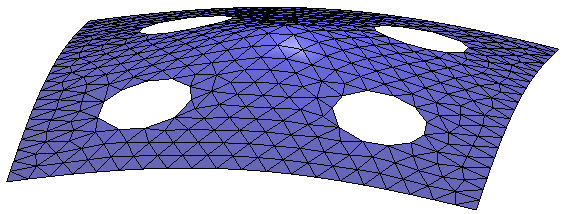}
		\includegraphics[width=0.3\textwidth]{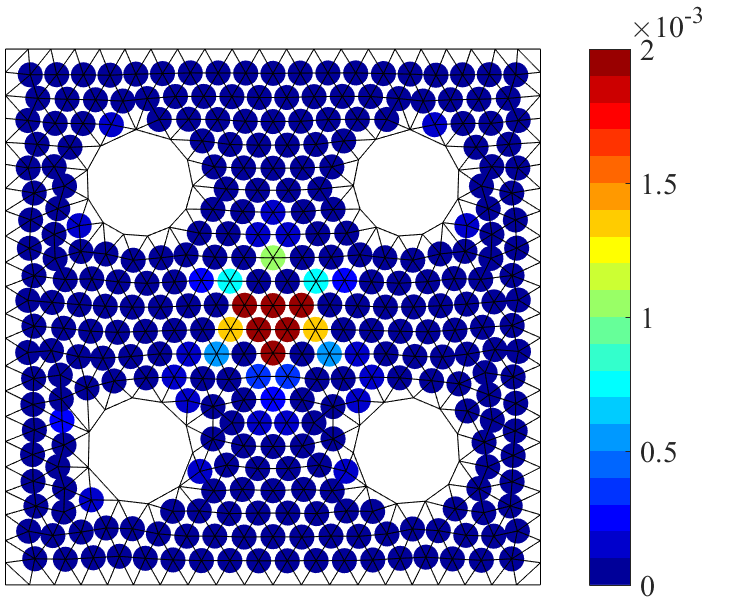}
	\caption{Initial settings of Example 4.
	From left to right: perspective view, Gaussian curvatures.}
	\label{Fig:Ex5}
\end{figure*}

In this example, we consider a surface with a non-trivial topology
having four circular openings.
The initial geometry, as shown in \reffig{Fig:Ex5}, is generated by linearly interpolating
the centre vertex and the opening.
It has almost zero Gaussian curvatures at all vertices except for the centre corn vertex,
where the Gaussian curvature concentrates.
Its span is 40.0m and its height is 8.0m.
This surface is composed of 438 vertices (of which 85 are boundary vertices), 1193 edges, and 752 faces.
We set $\mathcal{E}_{\text{fix}}$ to be all boundary edges including those of four openings,
while $\mathcal{V}_{\text{fix}}$ to be outer boundary vertices excluding those of four openings.
This means that the lengths of edges bounding the four openings are preserved but not their positions. We set $\lambda_e=0.01$ and $\lambda_v=0.001$.
The conformal structure $\eta_{ij}$ is calculated from the initial geometry by \eqref{Eq:inversive} and \eqref{Eq:eta}.
We have two variations of Example 4. 
For Example 4a, the target Gaussian curvature is set to 0.002 at the center and zero around the four corners, and linearly interpolated at other internal vertices.
On the other hand, 
for Example 4b, the target Gaussian curvature is set to zero at the center and 0.002 around the four corners, and linearly interpolated at other internal vertices.

\begin{figure*}[ht]
	\centering
	\hspace{-20mm}
	\includegraphics[width=0.65\textwidth]{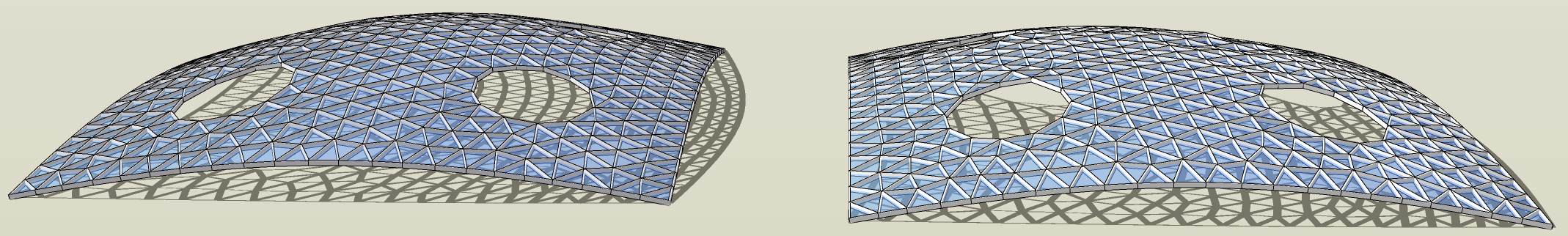}
		\includegraphics[width=0.22\textwidth]{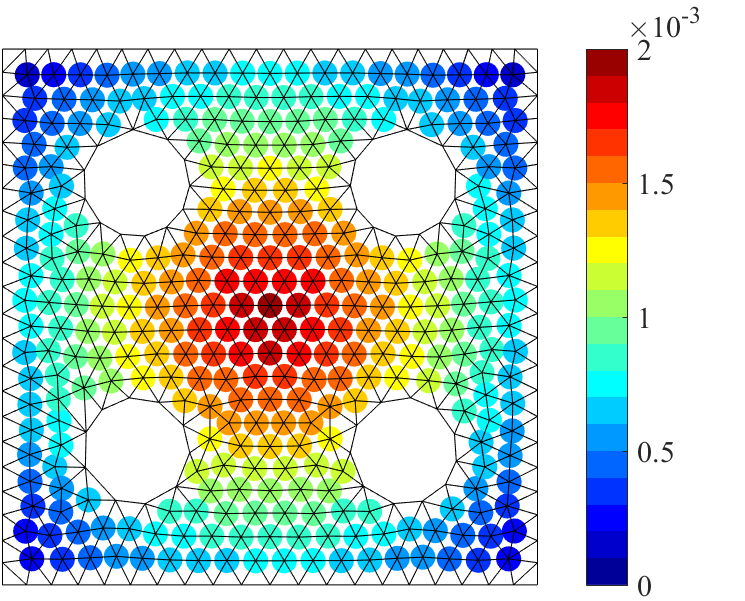}
		\includegraphics[width=0.22\textwidth]{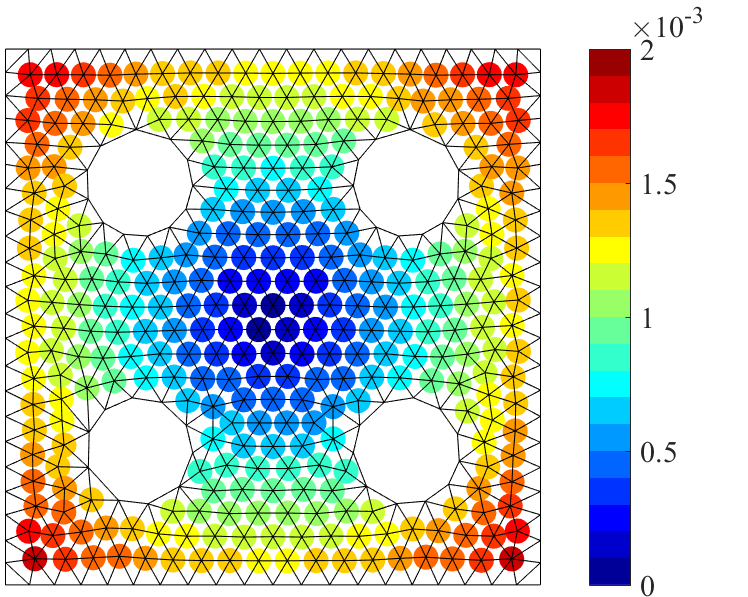}
	\caption{Final geometry of Example 4.
		From left to right: Example 4a, Example 4b,
		final Gaussian curvatures of Example 4a, Example 4b.
	Though the appearances of the two surfaces are very close,
	the structural stiffness of the left one is 1.7 times higher than the right one.}
	\label{Fig:Ex5_final}
\end{figure*}

The final geometries and the distributions of Gaussian curvatures
for these two examples are shown in  \reffig{Fig:Ex5_final}.
We observe that the final geometries are close to their initial geometry while they have the user-defined Gaussian curvatures.
{The two geometries look quite similar, however, they have different structural properties.
The pin-jointed analysis models of the structures with the fixed boundaries show that
the structural stiffness, evaluated by external works, of Example 4a is about 1.7 times higher than that of Example 4b.
This is mainly due to the lower stiffness (or larger displacements) in the middle part of Example 4b,
where the vertices have negative Gaussian curvatures.
}

\subsection*{Example 5: pointed dome with a side of various constant curvatures}
\begin{figure*}[ht]
	\centering
	\includegraphics[width=0.25\textwidth]{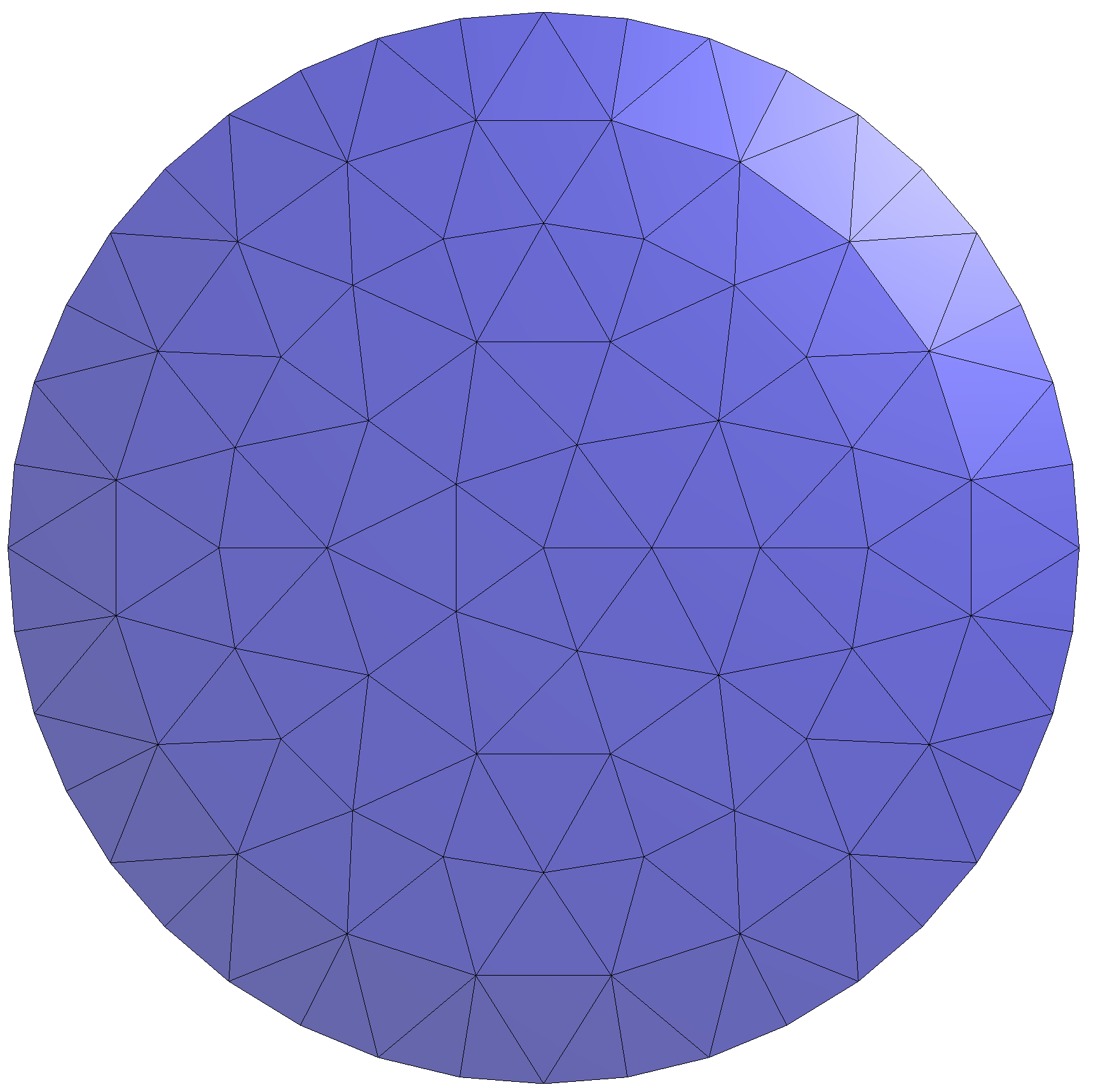}
	\includegraphics[width=0.35\textwidth]{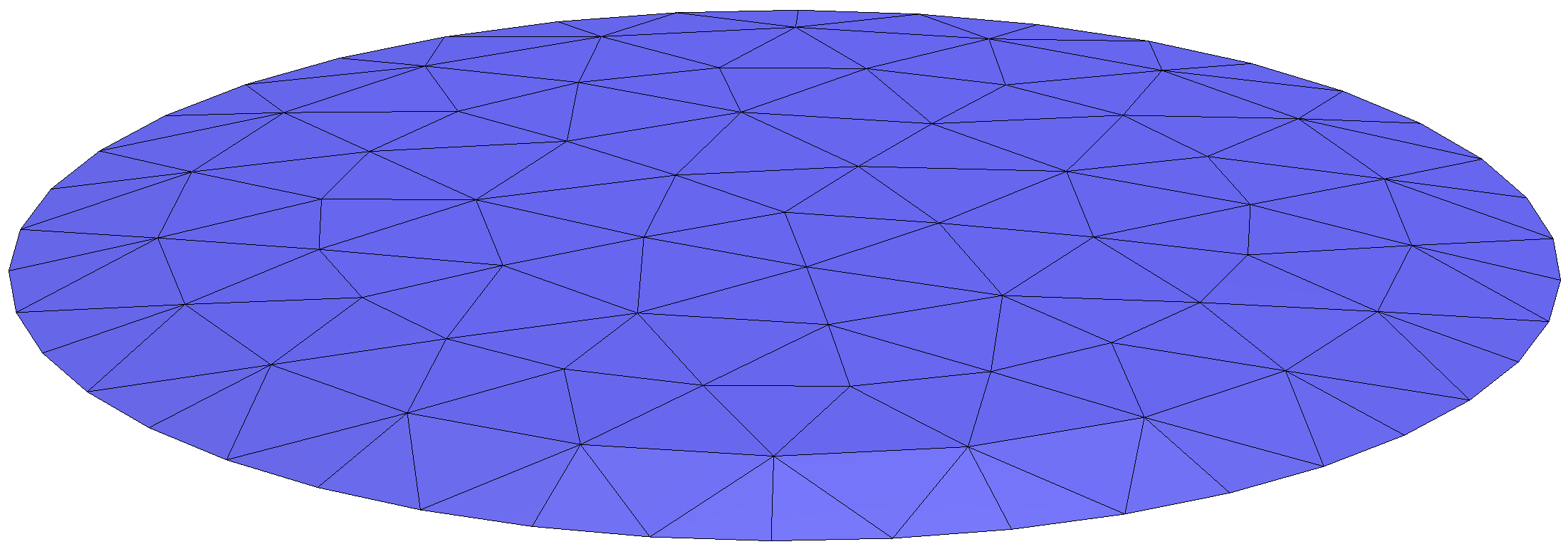}
	\includegraphics[width=0.3\textwidth]{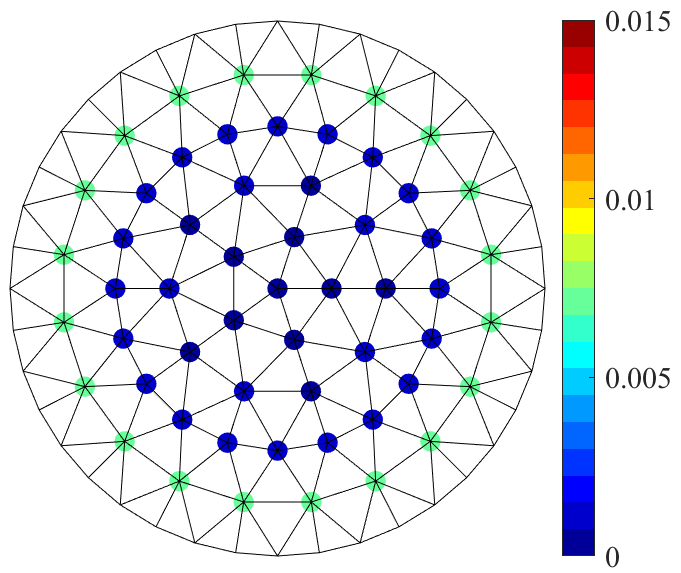}
	\caption{Initial settings of Example 5.
		From left to right: top view, perspective view, Gaussian curvatures.}
	\label{Fig:Ex4}
\end{figure*}

\begin{figure*}[ht]
	\centering
	\includegraphics[height=0.2\textwidth]{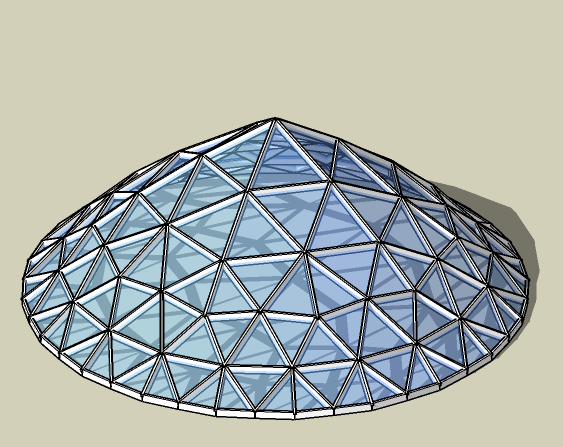}
	\includegraphics[height=0.2\textwidth]{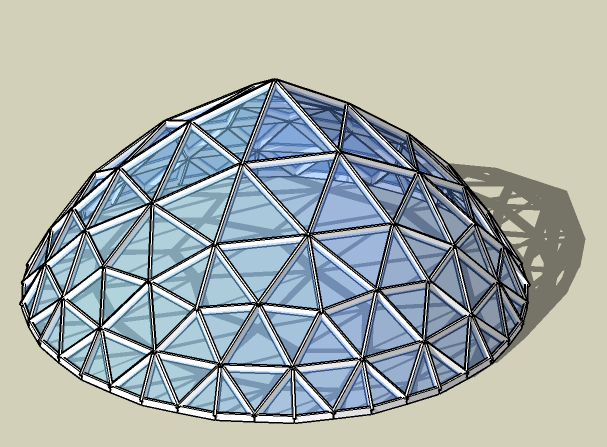}
	\includegraphics[height=0.2\textwidth]{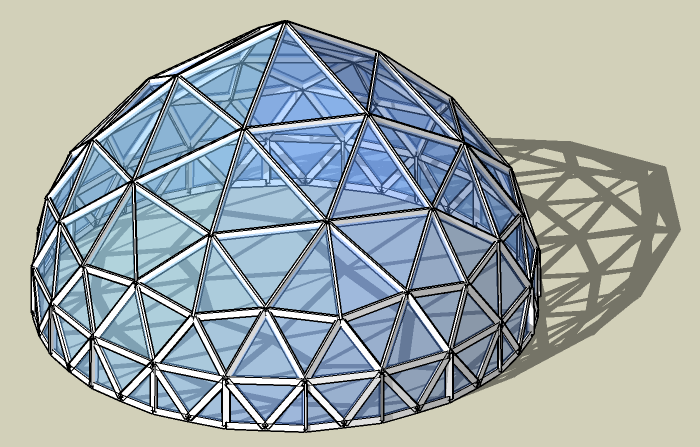}
	\caption{Renderings of the final design of Example 5.
		From left to right: Example 5a, Example 5b, Example 5c
		with a fixed Gaussian curvature at the apex and increasing Gaussian curvatures of the sides.}
	\label{Fig:Ex4_final}
\end{figure*}

In this example, our algorithm is used to find the design of domes, where
	the apex has a fixed curvature of $\pi/3$ and the side has the constant target curvature $K_0$.
	The initial design is given Figure \ref{Fig:Ex4}, which is almost flat 
	with 96 vertices and 150 faces.
	We set $\lambda_e=0.01$, $\lambda_c=0.1$, and $\lambda_v=0.001$.
	\reffig{Fig:Ex4_final} shows the final design with three different target curvatures $K_0\in \{0.02, 0.05, 0.1\}$, which give quite different silhouette and impression.
	The target curvature can be chosen interactively, providing an easy design tool.
	Note that if we do not add the convexity constraint and set $\lambda_c=0$,
	the resulting final design is concave, 
	demonstrating the effectiveness of the convexity term \eqref{Eq:convex}.

\subsection*{Example 6: interactive surface design}
\begin{figure}[!h]
		\centering
		\includegraphics[angle=90,origin=c,height=0.85\textheight]{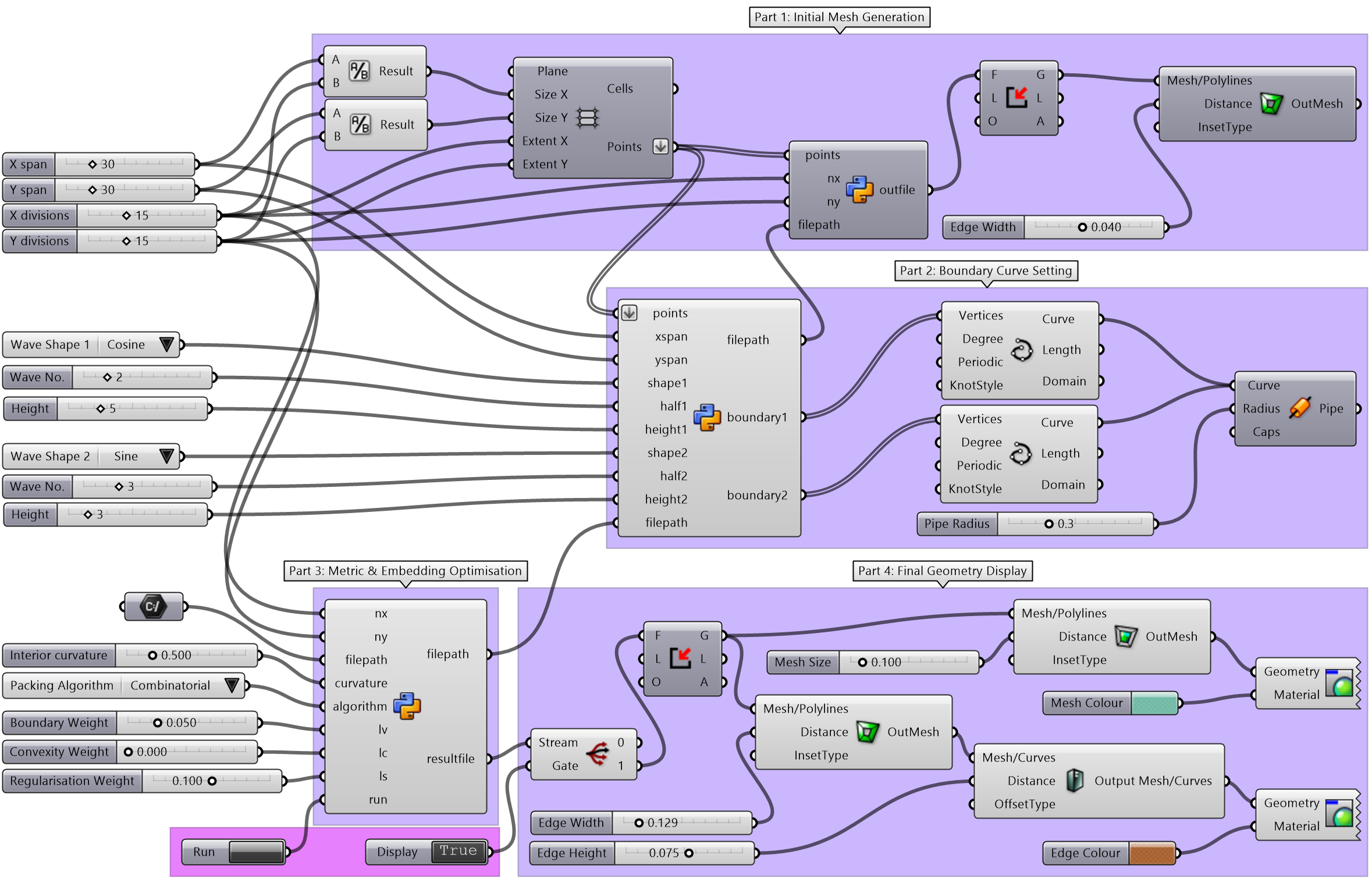}
		\caption{Interactive design tool integrated into Rinoceros 3D with Grasshopper.}
		\label{Fig:Ex6_Grasshopper2}
	\end{figure}

\begin{figure*}[ht]
	\centering
	\hspace{-20mm}
		\includegraphics[width=0.4\textwidth]{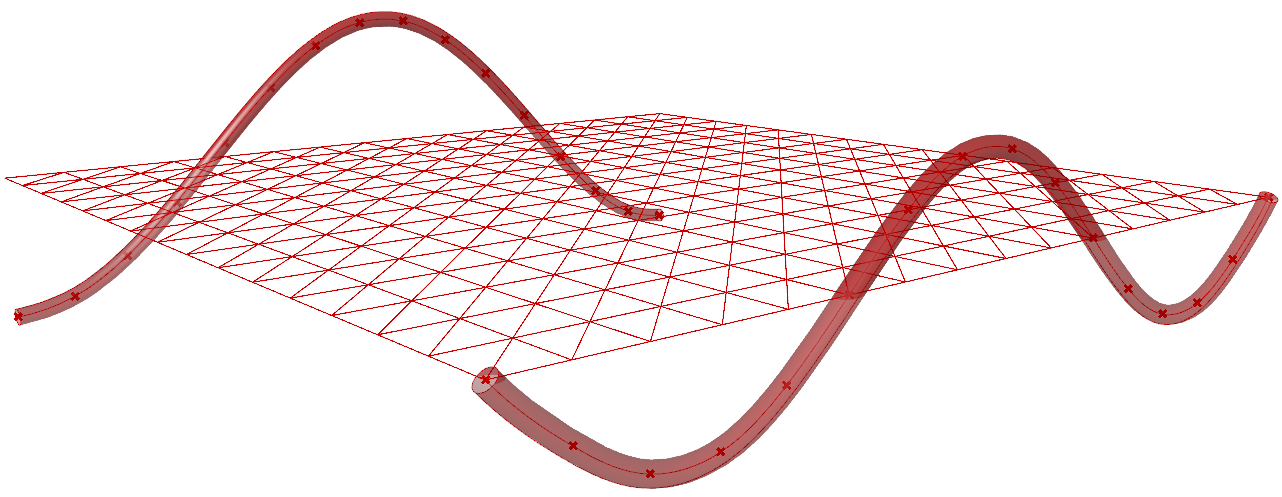}
		\includegraphics[width=0.35\textwidth]{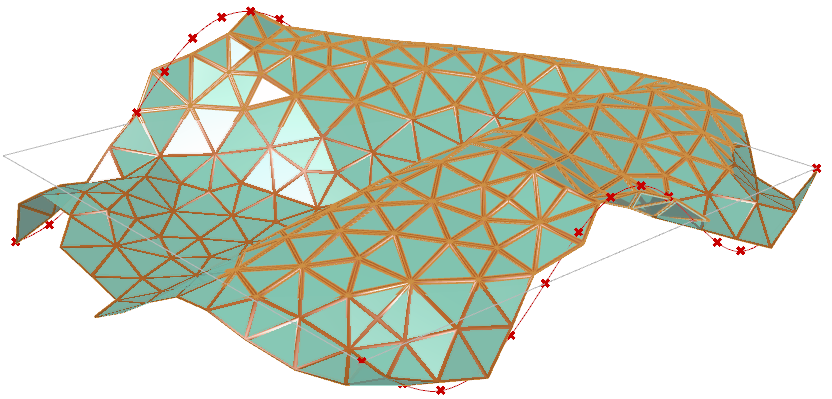}
	\includegraphics[width=0.30\textwidth]{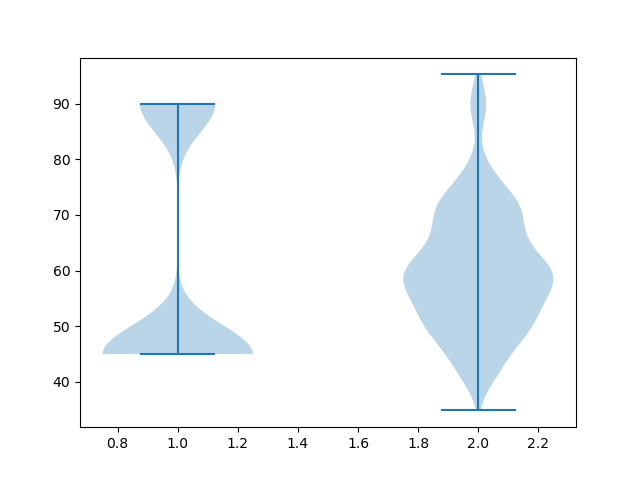}
	\caption{The design obtained by the interactive tool in Example 6.
	Left to right: specification of the boundary curves and the mesh topology,
	the resulting geometry,
	and the distributions of the initial (left) and the final (right) corner angles.
}
	\label{Fig:Ex6}
\end{figure*}

This subsection demonstrates the interactive tool 
integrated to the widely-used CAD software, Rhinoceros 3D,
via its visual programming language Grasshopper.
Rhinoceros 3D and Grasshopper are particularly suitable for parametric design of architecture and fabrication~\cite{Tedeschi2011}. 
Our design tool comprises the following four parts as shown in \reffig{Fig:Ex6_Grasshopper2}: 
\begin{itemize}
	\item \textbf{Part 1: Initial Mesh Generation}\\
		The initial geometry lying on the $xy$-plane is generated according to the user specified
		spans in $x$- and $y$-directions for a square plan,
		and the divisions in each direction.
	\item \textbf{Part 2: Boundary Curve Setting}\\
		Boundary curves are defined.
	\item \textbf{Part 3: Metric \& Embedding Optimisation}\\
    The metric optimization is performed with specified parameters.
    Two common conformal structure presets are defined;  
    ``Inversive'' to obtain panel shapes close to the initial one or ``Combinatorial'' 
    to obtain panel shapes close to equilateral triangles.
		The embedding optimisation is subsequently performed with the assigned weights.
	\item \textbf{Part 4: Final Geometry Display}\\
		The final geometry is displayed with simple rendering in the Rhinoceros 3D.
\end{itemize}
As an illustrative example, we create a CGC surface with a square plan with
the spans in both directions set to 30.0m 
(see the supplementary video).
The design region is uniformly divided into 15 sections in $x$- and $y$-directions. 
There are in total 256 vertices, including 196 inner vertices and 60 boundary vertices, 1410 edges, and 450 triangle panels.
One of the boundary curve has the shape of two half cosine waves with the height is 5.0m,
and the other has the shape of three half sine waves with the height of 3.0m,
as indicated by the thicker curves in \reffig{Fig:Ex6} (left figure).
The coordinates of the 32 boundary vertices lying on the sine and cosine curves 
are assigned in Part 2 of the design tool, 
and the coordinates of the other 24 boundary vertices are automatically determined by the tool.
The total target Gaussian curvature for interior vertices is set to $0.5$ (``Interior curvature''),
and hence, the target Gaussian curvature ${\bar K}_i$ of each interior vertex is 2.551$\times 10^{-3}$.
For the conformal structure, 
we set the constant value $\eta_{ij}=1$ by choosing ``Combinatorial'' in Part 3, 
aiming at equilateral panels.
The final geometry shown in \reffig{Fig:Ex6} is obtained by
the embedding optimisation with the parameters $\lambda_v=0.05$ (``Boundary Weight''), 
$\lambda_c=0.0$ (``Convexity Weight''), $\lambda_r=0.1$ (``Regularization Weight'').
The distributions of the Gaussian curvatures and the corner angles at the interior vertices for the initial and final meshes
are also shown in \reffig{Fig:Ex6}.
It is notable that the specified Gaussian curvatures, the boundary, and the corner angles cannot be achieved at the
same time, and trade-off among these factors has be to made
and the user can set the priority through the choice of hyper-parameters.

\section{Conclusions}\label{sec:conclusion}


In this study, we presented an efficient tool for the design of 
 discrete architectural surfaces having the
user-specified Gaussian curvatures and boundary locations while providing
control over the shape of the panels through the choice of the conformal class.
The proposed method is composed of two successive optimisation procedures:
the metric optimisation using a modified Ricci energy
and the embedding optimisation to realise the obtained metric.
Both are formulated as least-squares problems, and any standard solvers can be used
thanks to the closed-form expression of the modified Ricci energy.
This formulation and the ``change of the variables'' in the optimisation problems
are the key ideas of the proposed method.
Instead of taking vertex coordinates directly as variables,
the circle packing radii are chosen first as variables to find a metric.
Then, vertex coordinates are used as variables to achieve the obtained metric.
These ideas have led to our fast and robust algorithm.
In addition, our method offers flexibility in incorporating various design constraints as penalty terms
in the least-squares problem.
Our implementation is publicly available as open-source 
and can be easily extended with 
additional penalty terms and different solvers.

{One limitation of the proposed method is that only 
triangulated mesh can be handled 
due to the lack of the theory of conformal geometry for general polyhedral meshes
or parametric surfaces such as NURBS surfaces.
Another limitation is the second step of our optimisation, which is sensitive to the initial geometry.
Iterating the first and second steps alternatingly may be required to achieve a plausible balance
between the two optimisation steps.
Preferably, a more sophisticated method to incorporate the second step into the first one should be developed in the future.
}

\bigskip
\noindent\textbf{Acknowledgement:}
This work was supported by JST CREST Grant Number JPMJCR1911.
The authors are grateful to Prof. Makoto Ohsaki at Kyoto University for his helpful comments.

\bigskip
\bibliographystyle{plain}
\bibliography{references}



\end{document}